%% file: final_draft.tex
\input harvmac
\input epsf
\input amssym
%\draftmode
\baselineskip 13pt

\input youngtab.tex
\font\small=cmss10 at 4pt

%\pdfpageheight 0pt
%\pdfpagewidth 0pt

\def\IZ{\Bbb{Z}}
\def\IC{\Bbb{C}}

%%%%%%%%%ERIC's DEFS:

%%%%%% NEW DEFS:

\def\L{\Lambda}

\def\bb{{\bf b}}
\def\bo{{\bf \omega}}
\def\be{{\bf e}}
\def\br{{\bf{\rho}}}

\def\L{\Lambda}

\def\hsl{hs[$\lambda$]}

\def\ad{a^{\dagger}}
\def\ah{\widehat{a}}

\def\hsh{hs$\left[{1\over 2}\right]$}

\def\p{\partial}

\def\Oc{{\cal O}}
\def\half{{1\over 2}}
\def\rar{\rightarrow}

\def\a{\alpha}
\def\b{\beta}
\def\o{\omega}
\def\l{\lambda}

\def\Ab{\overline{A}}

\def\ab{\overline{a}}

\def\zb{\overline{z}}

\def\abar{\overline{a}}

\def\vs{\vskip .1 in}
\def\bul{$\bullet$~}

\def\IZ{\Bbb{Z}}

%%%%%%%%%%%%%%%%%%%%%%%%%%%%%%%%%%%%%%%%%%%%%%
%%%%%%%%%%%%%%%%%%%%%%%%%%%
% some stuff needed for figures:
%%%%%%%%%%%%%%%%%%%%%%%%%%%%%%%%%%%%%%%%%%%%%%
%%%%%%%%%%%%%%%%%%%%%%%%%%%
\newcount\figno
\figno=0
\def\fig#1#2#3{
\par\begingroup\parindent=0pt\leftskip=1cm\rightskip=1cm\parindent=0pt
\baselineskip=11pt
\global\advance\figno by 1
\midinsert
\epsfxsize=#3
\centerline{\epsfbox{#2}}
\vskip -21pt
{\bf Fig.\ \the\figno: } #1\par
\endinsert\endgroup\par
}
\def\figlabel#1{\xdef#1{\the\figno}}
\def\encadremath#1{\vbox{\hrule\hbox{\vrule\kern8pt\vbox{\kern8pt
\hbox{$\displaystyle #1$}\kern8pt}
\kern8pt\vrule}\hrule}}
%%%%%%%%%%%%%%%%%%%%%%%%%%%%%%%%%%%%%%%%%%%%%%

\def\p{\partial}

\def\rt{\rightarrow}
\def\Oc{{\cal O}}

\def\av{\vec{a}}

\def\zb{\overline{z}}

\def\abar{\overline{a}}

\def\Ab{\overline{A}}

\def\zb{\overline{z}}

\def\ab{\overline{a}}

\def\cb{\overline{c}}

\def\gb{\overline{g}}

\def\wb{\overline{w}}

\def\Lamb{\overline{\Lambda}}

\def\gb{\overline{g}}

\def\IZ{\Bbb{Z}}
\def\hw{{\rm hw}}

%\KrausUF
\lref\KrausUF{
  P.~Kraus and E.~Perlmutter,
  ``Probing higher spin black holes,''
[arXiv:1209.4937 [hep-th]].
%%CITATION = arXiv:1209.4937%%
}

%\GaberdielUJ
\lref\GaberdielUJ{
  M.~R.~Gaberdiel and R.~Gopakumar,
  ``Minimal Model Holography,''
[arXiv:1207.6697 [hep-th]].
%%CITATION = arXiv:1207.6697%%
}

%\PR
\lref\PR{
  K.~Papadodimas and S.~Raju,
  ``Correlation Functions in Holographic Minimal Models,''
Nucl.\ Phys.\ B {\bf 856}, 607 (2012).
[arXiv:1108.3077 [hep-th]].
%%CITATION = arXiv:1108.3077%%
}

%\CastroIW
\lref\CastroIW{
  A.~Castro, R.~Gopakumar, M.~Gutperle and J.~Raeymaekers,
  ``Conical Defects in Higher Spin Theories,''
JHEP {\bf 1202}, 096 (2012).

[arXiv:1111.3381 [hep-th]].
%%CITATION = arXiv:1111.3381%%
}

%\ChangYin
\lref\ChangYin{
 C-M.~Chang and X.~Yin,
  ``Correlators in $W_N$ Minimal Model Revisited,''
JHEP {\bf 1007},10(2012)050. [arXiv:1112.5459 [hep-th]].
%%CITATION = arXiv:1112.5459%%
}

%\PopeSR
\lref\PopeSR{
  C.~N.~Pope, L.~J.~Romans and X.~Shen,
  ``W(infinity) And The Racah-wigner Algebra,''
Nucl.\ Phys.\ B {\bf 339}, 191 (1990)..
%%CITATION = USC-89-HEP040%%
}

%\GaberdielZW
\lref\GaberdielZW{
  M.~R.~Gaberdiel, R.~Gopakumar, T.~Hartman and S.~Raju,
  ``Partition Functions of Holographic Minimal Models,''
JHEP {\bf 1108}, 077 (2011).
[arXiv:1106.1897 [hep-th]].
%%CITATION = arXiv:1106.1897%%
}

%\GaberdielKU
\lref\GaberdielKU{
  M.~R.~Gaberdiel and R.~Gopakumar,
  ``Triality in Minimal Model Holography,''
[arXiv:1205.2472 [hep-th]].
%%CITATION = arXiv:1205.2472%%
}

%\ProkushkinBQ
\lref\ProkushkinBQ{
  S.~F.~Prokushkin and M.~A.~Vasiliev,
  ``Higher spin gauge interactions for massive matter fields in 3-D AdS space-time,''
Nucl.\ Phys.\ B {\bf 545}, 385 (1999).
[hep-th/9806236].
%%CITATION = hep-th/9806236%%
}

%\GaberdielPZ
\lref\GaberdielPZ{
  M.~R.~Gaberdiel and R.~Gopakumar,
  ``An $AdS_3$ Dual for Minimal Model CFTs,''
Phys.\ Rev.\ D {\bf 83}, 066007 (2011).
[arXiv:1011.2986 [hep-th]].
%%CITATION = arXiv:1011.2986%%
}

%\BlencoweGJ
\lref\BlencoweGJ{
  M.~P.~Blencowe,
  ``A Consistent Interacting Massless Higher Spin Field Theory In D = (2+1),''
  Class.\ Quant.\ Grav.\  {\bf 6}, 443 (1989).
  %%CITATION = CQGRD,6,443;%%
}

%\GaberdielWB
\lref\GaberdielWB{
  M.~R.~Gaberdiel and T.~Hartman,
  ``Symmetries of Holographic Minimal Models,''
  arXiv:1101.2910 [hep-th].
  %%CITATION = ARXIV:1101.2910;%%
}

%\KhesinEY
\lref\KhesinEY{
  B.~Khesin and F.~Malikov,
  ``Universal Drinfeld-Sokolov reduction and matrices of complex size,''
Commun.\ Math.\ Phys.\  {\bf 175}, 113 (1996).
[hep-th/9405116].
%%CITATION = hep-th/9405116%%
}

%\KrausDS
\lref\KrausDS{
  P.~Kraus and E.~Perlmutter,
  ``Partition functions of higher spin black holes and their CFT duals,''
JHEP {\bf 1111}, 061 (2011).
[arXiv:1108.2567 [hep-th]].
%%CITATION = arXiv:1108.2567%%
}

%\HenneauxXG
\lref\HenneauxXG{
  M.~Henneaux and S.~J.~Rey,
  ``Nonlinear W(infinity) Algebra as Asymptotic Symmetry of Three-Dimensional
  Higher Spin Anti-de Sitter Gravity,''
  JHEP {\bf 1012}, 007 (2010)
  [arXiv:1008.4579 [hep-th]].
  %%CITATION = JHEPA,1012,007;%%
}

%\ChangKT
\lref\ChangKT{
  C.~-M.~Chang, S.~Minwalla, T.~Sharma and X.~Yin,
  ``ABJ Triality: from Higher Spin Fields to Strings,''
[arXiv:1207.4485 [hep-th]].
%%CITATION = arXiv:1207.4485%%
}
%\CampoleoniZQ
\lref\CampoleoniZQ{
  A.~Campoleoni, S.~Fredenhagen, S.~Pfenninger and S.~Theisen,
  ``Asymptotic symmetries of three-dimensional gravity coupled to higher-spin
  fields,''
  JHEP {\bf 1011}, 007 (2010)
  [arXiv:1008.4744 [hep-th]].
  %%CITATION = JHEPA,1011,007;%%
}

%\AmmonUA
\lref\AmmonUA{
  M.~Ammon, P.~Kraus and E.~Perlmutter,
  ``Scalar fields and three-point functions in D=3 higher spin gravity,''
[arXiv:1111.3926 [hep-th]].
%%CITATION = arXiv:1111.3926%%
}

%\ChangMZ
\lref\ChangMZ{
  C.~-M.~Chang and X.~Yin,
  ``Higher Spin Gravity with Matter in $AdS_3$ and Its CFT Dual,''
[arXiv:1106.2580 [hep-th]].
%%CITATION = arXiv:1106.2580%%
}

%\FateevAB
\lref\FateevAB{
  V.~A.~Fateev and A.~V.~Litvinov,
  ``Correlation functions in conformal Toda field theory. I.,''
JHEP {\bf 0711}, 002 (2007).
[arXiv:0709.3806 [hep-th]].
%%CITATION = arXiv:0709.3806%%
}

%\GutperleKF
\lref\GutperleKF{
  M.~Gutperle and P.~Kraus,
  ``Higher Spin Black Holes,''
JHEP {\bf 1105}, 022 (2011).
[arXiv:1103.4304 [hep-th]].
%%CITATION = arXiv:1103.4304%%
}

%\ChangIZP
\lref\ChangIZP{
  C.~-M.~Chang and X.~Yin,
  ``A semi-local holographic minimal model,''
[arXiv:1302.4420 [hep-th]].
%%CITATION = arXiv:1302.4420%%
}

%\JevickiKMA
\lref\JevickiKMA{
  A.~Jevicki and J.~Yoon,
  ``Field Theory of Primaries in $W_N$ Minimal Models,''
[arXiv:1302.3851 [hep-th]].
%%CITATION = arXiv:1302.3851%%
}

%\DiFrancescoNK
\lref\DiFrancescoNK{
  P.~Di Francesco, P.~Mathieu and D.~Senechal,
  ``Conformal field theory,''
New York, USA: Springer (1997) 890 p.
}

%\PerlmutterDS
\lref\PerlmutterDS{
  E.~Perlmutter, T.~Prochazka and J.~Raeymaekers,
  ``The semiclassical limit of $W_N$ CFTs and Vasiliev theory,''
[arXiv:1210.8452 [hep-th]].
%%CITATION = arXiv:1210.8452%%
}
%\WittenUA
\lref\WittenUA{
  E.~Witten,
  ``Multitrace operators, boundary conditions, and AdS / CFT correspondence,''
[hep-th/0112258].
%%CITATION = hep-th/0112258%%
}

%\ElShowkAG
\lref\ElShowkAG{
  S.~El-Showk and K.~Papadodimas,
  ``Emergent Spacetime and Holographic CFTs,''
JHEP {\bf 1210}, 106 (2012).
[arXiv:1101.4163 [hep-th]].
%%CITATION = arXiv:1101.4163%%
}

%\HeemskerkPN
\lref\HeemskerkPN{
  I.~Heemskerk, J.~Penedones, J.~Polchinski and J.~Sully,
  ``Holography from Conformal Field Theory,''
JHEP {\bf 0910}, 079 (2009).
[arXiv:0907.0151 [hep-th]].
%%CITATION = arXiv:0907.0151%%
}

%\DotsenkoF
\lref\DotsenkoF{
  V.~S. ~Dotsenko and V.~A. ~Fateev,
  ``Conformal algebra and multipoint correlation functions in 2D statistical models,''
Nucl. Phys. B {\bf 240}, (1984) 312; ibid. {\bf 251} (1985) 691
}

%\BlaisBS
\lref\BlaisBS{
 F.~A. ~Blais, P. ~Bouwknegt and M. ~Surridge,
  ``Coset construction for extended Virasoro algebras,''
Nucl. Phys. B {\bf 304}, (1988) 371
}

%\GiombiMS
\lref\GiombiMS{
  S.~Giombi and X.~Yin,
  ``The Higher Spin/Vector Model Duality,''
[arXiv:1208.4036 [hep-th]].
%%CITATION = arXiv:1208.4036%%
}

%\BerkoozUG
\lref\BerkoozUG{
  M.~Berkooz, A.~Sever and A.~Shomer,
  ``'Double trace' deformations, boundary conditions and space-time singularities,''
JHEP {\bf 0205}, 034 (2002).
[hep-th/0112264].
%%CITAT
}
%%%%%%%%%%%%%%%%%%%%%%%

\Title{\vbox{\baselineskip14pt
%\hbox{hep-th/0508218}
%\hbox{UCLA-05-TEP-XX} \hbox{MCTP-XX-XX}
}} {\vbox{\centerline {Matching four-point functions}
\medskip\vbox{\centerline {in higher spin AdS$_3$/CFT$_2$}}}}
\centerline{Eliot Hijano$^{\dag}$, Per Kraus$^{\dag}$ and Eric
Perlmutter$^{\star}$\foot{eliothijano@physics.ucla.edu,
pkraus@ucla.edu, E.Perlmutter@damtp.cam.ac.uk}}
\bigskip
\centerline{\it{$\dag$~Department of Physics and Astronomy}}
\centerline{${}$\it{University of California, Los Angeles, CA 90095, USA}}
\vs
\centerline{\it{$\star$~DAMTP, Centre for Mathematical Sciences,  University of Cambridge}}
\centerline{${}$\it{Cambridge, CB3 0WA, UK}}
\baselineskip14pt

\vskip .3in

\centerline{\bf Abstract}

Working in the context of  the proposed duality between 3D higher spin gravity and 2D $W_N$ minimal model CFTs, we compute a class of four-point functions in the bulk and on the boundary, and demonstrate precise agreement between them. This is the first computation of a correlator in 3D higher spin gravity whose functional form is not fixed by conformal invariance.  In the bulk we make use of elegant methods to solve the scalar field equation using only matrix manipulations, while on the CFT side we employ the Coulomb gas representation.   Comparison is made in the semiclassical  limit, in which the central charge is taken to infinity at fixed $N$. Along the way, we  establish the rules for computing correlation functions of multi-trace operators in higher spin gravity.  The method involves  solving the scalar master field equation in a general representation of the bulk gauge algebra.  Although most of our work is carried out for the bulk theory based on sl(N) $\times$ sl(N), we also discuss how our methods can be adapted to the  \hsl\ $\times$ \hsl\ theory, which is relevant in the 't Hooft limit.

%%%
\Date{February  2013}
%%%%%%%%%%%%%%%%%%%%%%%%%%%%%%%%%%%%%%%%%%%%%%
%%%%%%%%%%%%%%%%%%%%%%%%%%%
% Main text begins here
%%%%%%%%%%%%%%%%%%%%%%%%%%%%%%%%%%%%%%%%%%%%%%
%%%%%%%%%%%%%%%%%%%%%%%%%%%
\baselineskip13pt

\listtoc\writetoc

\newsec{Introduction}

A promising approach to gaining insight into the mechanism underlying holography is the study of dualities involving higher spin gravity theories in AdS; see \refs{\GaberdielUJ,\GiombiMS} for recent reviews.   The boundary CFTs in these cases are exactly solvable, allowing us to examine in detail how bulk and boundary physics are related, without being confined to the straitjacket of considering only BPS protected quantities.   In this paper we focus on the  proposal of Gaberdiel and Gopakumar, which relates three-dimensional higher spin gravity to the $W_N$ minimal models \refs{\GaberdielPZ,\GaberdielKU}.  These CFTs can be thought of as the natural generalization of the Virasoro minimal models and, like the latter, have a coset description,
\eqn\aa{ {SU(N)_k \oplus SU(N)_1 \over SU(N)_{k+1}}~.}
These are nontrivial, interacting  CFTs, yet they are exactly solvable: the full spectrum of allowed operator dimensions is known, and arbitrary correlation functions can in principle be computed.   The main reasons why these CFTs naturally appear on the boundary side of higher spin AdS/CFT dualities are: 1) their symmetry algebras are higher spin extensions of the Virasoro algebra, matching the asymptotic symmetries of the bulk theory \refs{\CampoleoniZQ,\HenneauxXG,\GaberdielWB}; 2) these theories admit a limit in which the central charge becomes parametrically large, corresponding to a classical limit in the bulk;  3) in such a limit, the CFT central charge can exhibit vector-like rather than adjoint scaling behavior, which is a characteristic of the spectrum of the standard Vasiliev higher spin  theory \ProkushkinBQ.

There are different ways to take the large central charge limit, as we discuss
more thoroughly in Section 2.   One is the 't Hooft limit, advocated
in the original paper \GaberdielPZ.   Here $N$ and $k$ are taken to
infinity, holding fixed $\lambda = N/(N+k)$.   The main puzzle that
emerges in this limit is the appearance of ``light states". This
label refers to an infinite class of primary operators whose
conformal dimensions scale as $1/N$, vanishing in the 't Hooft
limit.  Such states are not present in the perturbative spectrum of
the standard Vasiliev  theory in the bulk, yet they interact with ``ordinary sector'' states at leading order in $1/N$ on the CFT side \refs{\PR,\ChangYin}. So
apparently one must either identify these states as exotic
nonperturbative objects, or augment the bulk spectrum with
additional degrees of freedom.   A second option, which we refer to
as the ``semiclassical limit", is to hold $N$ fixed while taking
$\lambda =-N$, with $\lambda$ defined as above \GaberdielKU.    In this case, the
full spectrum of CFT primaries matches up nicely to states in the
bulk, without the need for additional degrees of freedom \refs{\GaberdielKU,\CastroIW, \PerlmutterDS}.   However,
one's eyebrows should be raised by the fact that setting $\lambda
=-N$ means that $k$ is taken to be negative, which renders the CFT
non-unitary; for instance, operators acquire negative dimension.

In this work we focus on exploring the semiclassical limit. The bulk
dual is conjectured to be the classical 3D Vasiliev theory \ProkushkinBQ\ at the specific point $\l=-N$, where we recall that $\l$ in the bulk is a free parameter characterizing the higher spin symmetry algebra.
For the considerations relevant to this paper, we can think of this theory in terms of flat sl(N) $\times$ sl(N)  Chern-Simons connections coupled to scalar matter in a
gauge invariant manner.  Although the
non-unitary nature of the theory will be manifest throughout, this
will not pose any obstacle towards verifying precise and detailed agreement
between bulk and boundary observables in the limit of large central
charge.  At higher orders in the $1/c$ expansion quantum effects in
the bulk will be important, and here the  lack of unitarity will
have to be confronted.  We postpone this issue for the future.

The AdS/CFT dictionary allows us to compare correlation functions computed in the bulk and boundary descriptions.  Let us recall which correlators have been successfully matched so far.  First, the fact that the symmetry algebras of the bulk and boundary theories agree \refs{\HenneauxXG, \CampoleoniZQ, \GaberdielWB} guarantees that correlators involving just the higher spin currents will match, since these are fixed by the higher spin symmetry.  Second, a class of three-point functions, $\langle \overline{\cal O} {\cal O} J^{(s)}\rangle$, has been shown to match \refs{\ChangMZ,\AmmonUA}, where ${\cal O}$ denotes a primary operator dual to a perturbative scalar field in the bulk, and $J^{(s)}$ is the spin-$s$ current.  Of course, conformal invariance fixes the functional form of three-point functions, and so this match is just a check of the overall normalization.

Our main goal in this paper is to extend this story to the matching
of four-point functions.   Conformal invariance only fixes the
functional form of the four-point function  up to an arbitrary
function of the conformal cross ratio, and so matching this object
is a much more comprehensive check of the duality: it probes the detailed
structure of the specific CFT in consideration. In the semiclassical
limit we are able to compute a large class of four-point functions
with relative ease, as we now explain.  Primary operators in the CFT
are labelled as $(\Lambda_+,\Lambda_-)$, where $\Lambda_\pm$ denote
representations of SU(N).   In the semiclassical limit
$(\Lambda_+,0)$ operators map to perturbative scalar fields in the
bulk, while $(0,\Lambda_-)$ map to classical solitonic solutions \PerlmutterDS. We
refer to these as scalar and defect operators respectively.   The
four-point functions computed in this paper involve two scalar
operators and two defect operators.  The bulk computation of these
correlators can be recast in terms of the scalar two-point function
computed in the background of the soliton.  The soliton solutions
are built purely out of the higher spin gauge fields, and so are
described by certain flat connections.  We proceed by drawing on
elegant methods for solving the scalar wave equation in the
background of higher spin gauge fields, techniques that were
employed, for example, in the computation of scalar correlators in
higher spin black hole backgrounds \KrausUF.   The computation turns out to
involve only fairly elementary operations on $N\times N$ matrices.

On the CFT side, an efficient method for attacking correlators in coset models is the Coulomb gas formalism \refs{\DotsenkoF}.   Here one represents primary operators in terms of free bosons, along with the insertion of certain screening operators.   This procedure typically returns expressions for correlators in terms of hypergeometric functions and a sum over conformal blocks. Certain correlators were computed using this method in the $W_N$ holography context in \refs{\PR,\ChangYin}.

One of our main results is the matching of the four-point functions in the case that the scalar operator is in the defining representation while the defect representation is kept arbitrary.  We first compute this on the bulk side, with $\lambda=-N$,  and find the surprisingly simple result (4.17). On the CFT side the corresponding result for general $k$ and $N$ is a more complicated expression in terms of a generalized hypergeometric function, (4.29), yet remarkably, upon taking the $\lambda \rightarrow -N $ limit the answer collapses down to display perfect agreement with the aforementioned bulk result.

The extension of the bulk computation to the case of an arbitrary scalar operator $(\Lambda_+,0)$ involves some novelties.   Our computation in the case that $\Lambda_+$ is the defining  representation involves solving the  equation $dC + AC - C \overline{A}=0$, where $C$ is the master field whose trace gives the physical scalar field.    $A$ and $\overline{A}$ denote the higher spin gauge connections expressed in the defining representation; in particular, $A$, $\overline{A}$ and $C$ are all $N\times N$ matrices.   To extend this to the case of general $\Lambda_+$ we propose to solve the same master field equation, but now with the connections expressed in the representation $\Lambda_+$.

This proposal passes certain elementary consistency checks and should apply to the calculation of scalar propagators quite generally. We also verify agreement with CFT four-point functions in one special case, in which $\Lambda_+ = {\small\yng(1,1)} $  and $\Lambda_-$ is the defining representation.  There is no fundamental obstacle to verifying this agreement for more general representations, but the computations become complicated.

An interesting feature of this proposal is that a master field $C$ in a general representation $\Lambda_+$ corresponds not to a single field in the bulk, but to a multiplet of fields with different masses.   The spectrum can be worked out by decomposing the sl(N) representation $\Lambda_+$ into irreducible representations of sl(2). The spectrum of sl(2) spins determines the spectrum of masses in AdS via the formula $m^2 =4 j(j+1)$.   When $\Lambda_+$ is the defining representation, only a single field is present, since we embed sl(2) in sl(N) such that the defining representation of sl(N) becomes the N-dimensional irreducible representation of sl(2).  The fact that fields of different masses are linked to one another by the symmetries of the higher spin theory is intriguing, and perhaps indicative of a connection to string theory and its tower of massive states.

Scalar operators for general $\Lambda_+$ are usually thought of as being analogous to multi-trace operators in gauge theories, and as being dual to multi-particle states in the bulk.    We have provided a prescription for the computation  of correlators involving these operators.  This is to be compared with how one incorporates multi-trace operators in ordinary AdS/CFT, namely by modifying the boundary conditions for the bulk scalar \refs{\WittenUA,\BerkoozUG}.  It would be interesting to  understand the relation between these approaches.  We also note that some support for our prescription for handling multiparticle states can be found in the expression for the scalar one-loop partition function of the bulk theory, which can be written as a sum over sl(N) characters \GaberdielZW.

While our main focus on the bulk side is the sl(N)$\times$ sl(N) theory, we also make some comments  on the case where the gauge group is hs[$\lambda$] $\times$ hs[$\lambda$].  We recall that hs[$\lambda$] is the infinite dimensional algebra that appears in the usual formulation of the Vasiliev theory.  It can be thought of as the analytic continuation of sl(N) to non-integer $N$.   Our treatment of scalar operators seems to carry over to this case.  On the other hand, the role of defect operators in this theory involves subtleties whose study we postpone to future work.

To actually compute scalar correlation functions at generic $\l$, we needed to introduce new techniques that generalize our sl(N) matrix computations and those of \KrausUF\ at $\l=1/2$. The main technical insight is the utility of an infinite-dimensional matrix representation of \hsl. These results stand somewhat on their own relative to the rest of the paper, and are likely to be useful for general computations in the \hsl\ theory going forward.

The remainder of this paper is organized as follows. In Section 2, we contrast the holographic dualities involving the 't Hooft and semiclassical limits of the $W_N$ minimal models. Section 3 reviews the construction of the conical defect solutions in the $\l=-N$ bulk theory, and the computation of scalar correlation functions in higher spin gravity. This sets the stage for Section 4 in which we present the bulk and boundary computations of CFT four-point functions, establishing agreement in the semiclassical limit for the case of a scalar field in the defining representation and the defect in an arbitrary representation. In Section 5, we present and test our proposal for bulk computations in the $\l=-N$ theory involving arbitrary $(\L_+,0)$ operators of the CFT. In Section 6, we apply this prescription to compute the CFT four-point function with the scalar in the
{\small\yng(1,1)} representation, and the defect in the defining
representation, from both bulk and boundary; once again these match in the semiclassical limit. Section 7 generalizes the results of Section 5 to generic $\l$, and along the way presents new technology for computing scalar bulk-boundary propagators for generic $\l$. Section 8 closes with some discussion, and two appendices contain supplementary material.

\newsec{The semiclassical and 't Hooft limits of $W_N$ minimal models}

Consider the $W_N$ minimal models, described by the coset \refs{\BlaisBS}
\eqn\ca{ {SU(N)_k \oplus SU(N)_1 \over SU(N)_{k+1} }}
with central charge
\eqn\cb{ c = (N-1) \Big(1- {N(N+1) \over (N+k)(N+k+1)}\Big)~.}
This theory has $W_N$ symmetry generated by higher spin currents of spin
$s=2,3,\ldots,N$. The spectrum of $W_N$ primary fields and their
descendants is parameterized by a pair of highest weights of
$SU(N)_k$ and $SU(N)_{k+1}$ as $(\L_+,\L_-)$, respectively; in hope that it is clear from the context, we will use $(\L_+,\L_-)$ to denote a given state, operator or representation. The holomorphic
conformal dimension of the primaries is
\eqn\cd{ h{(\Lambda_+,\Lambda_-)} = {1\over 2 p (p+1)}
\Big(\big|(p+1)(\Lambda_++\hat{\rho})-p(\Lambda_-+\hat{\rho})\big|^2
-\hat{\rho}^2 \Big) }
where $p=N+k$ and $\hat{\rho}$ is the Weyl vector of SU(N).

It is useful to exchange $k$ for $\l={N\over N+k}$ and write the central charge as
\eqn\ce{c = (N-1)(\l-1)(M-1)}
where
\eqn\cg{M = -{N\l\over N+\l}=-{N\over N+k+1}~.}
This form of $c$ is suggestive of the triality symmetry of the quantum $W_{\infty}[\l]$ algebra, $W^{qu}_{\infty}[N] \cong W^{qu}_{\infty}[\l] \cong W^{qu}_{\infty}[M]$, under which the central charge is by definition fixed \GaberdielKU.

There are multiple ways to take $c$ large. Consider holding $\l$ fixed; assuming $N$ and $k$ are both positive, we have $0\leq \lambda \leq 1$. In the 't Hooft limit one then takes $N\rar\infty$. In the semiclassical limit, one takes $M\rar-\infty$, which is to say, $\l \rar -N$ with $N$ fixed. For the time being, we take $N$ to be an integer greater than one, which forces $M$ to go off to negative rather than positive infinity. In each of these limits the central charge displays vector-like growth, linear in the large parameter.

The existence of distinct classical limits of a single quantum theory is characteristic of theories with strong-weak coupling dualities, and it may be useful to think of the relation between the 't Hooft and semiclassical limits in this spirit.  The map between these descriptions is not obvious, and an equivalence, should it exist, would only be seen after quantization.   Each limit, by virtue of having large central charge, is expected to have some classical gravity description, and the question of exactly what the dual theory is can {\it a priori} be expected to have different answers in each limit. As we now describe, this is clearly the case here, where only the semiclassical limit seems to match to a classical Vasiliev theory in its most basic form. The extent to which we understand the 't Hooft limit holographically is rather indirect in the above sense, and the dual theory in that limit may turn out to be an interesting if complicated extension of the usual Vasiliev theory. This is analogous to a general property of ordinary (i.e. non-higher spin) holography: while CFTs with a large $N$ expansion can be argued to have  holographic duals, only in special cases is the bulk theory governed by ordinary low energy gravity \refs{\HeemskerkPN, \ElShowkAG}.

The semiclassical limit of the CFT is dual to the Vasiliev theory at $\l=-N$, which upon consistently modding out all generators with spins $s>N$ becomes sl(N)~$\times$~sl(N) Chern-Simons theory with a higher spin gauge-invariant coupling to matter, and gravitational coupling $G_N \sim 1/c$.\foot{Actually, this statement is only true up to quadratic order in the scalar fields, which is all that will concern us in this paper.  The proper description of the theory at higher orders in the scalars is an interesting question for the future.} This duality is completely manifest on the level of the operator spectrum.  This is in stark contrast to the 't Hooft limit: strictly in the 't Hooft limit, the duality with the \hsl\ Vasiliev theory works well, but perturbation theory in $1/N$ reveals subtleties that are incompatible with the standard classical Vasiliev theory.

Let us briefly elaborate. A central issue is the status of states
$(\L_+,\L_-)$ with $\L_-\neq0$ in the 't Hooft limit, dubbed
``additional sector'' states in \PR. For simplicity, we focus on the
``light states'' which have $\L_+=\L_-$ and $O(1)$ boxes in the Young
diagram. Such states have vanishing conformal weight in the limit,
\eqn\ch{h(\L,\L) \approx {\l^2\over N^2}C_2(\L)}
where $C_2(\L) = \half\L\cdot (\L+2\hat{\rho})$ is the quadratic Casimir
of SU(N), and $\cdot$ denotes  an inner product on the
weight space. There are many such states. While they decouple from
the theory at strictly infinite $N$, any bulk dual to this theory
which captures the dynamics of the $1/N$ expansion must accommodate
these degrees of freedom; this is not a feature of the Vasiliev
theory of \ProkushkinBQ. Similar reasoning extends to all additional sector
states which have $\L_-\neq 0$, whether $\L_+=\L_-$ or not.\foot{A
study of three-point function coefficients involving additional
sector operators permits an enumeration of precisely what bulk
fields one appears to need in order to reproduce CFT correlation
functions; there is, of course, an infinite tower of such fields \refs{\JevickiKMA,\ChangIZP}.
See \ChangIZP\ for a recent  suggestion on the incorporation of these
fields into a consistent gravity theory.}

Unlike the additional sector states, the states $(\L_+,0)$ are under
control in the 't Hooft limit of the duality: these states form a
closed (but not modular-invariant) subsector of the CFT, and  are
visible in the bulk as perturbative scalar excitations. The lowest
such primary, $({\small\yng(1)},0)$, has
$h({\small\yng(1)},0) = {1+\l\over 2}$. It is dual to the
bulk scalar with $m^2=-1+\l^2$ and it exhibits generalized
free-field behavior in its correlation functions at large $N$.

So
while the 't Hooft limit is quite appealing on the CFT side as a
unitary limit of the minimal models, the nature of nearly the entire
spectrum and the CFT fusion rules seem to require exotic
modifications of the bulk theory.

On the other hand, the semiclassical limit displays a match between the full CFT spectrum and bulk states, free of the need for a hidden sector. The match, first conjectured in \GaberdielKU\ and then modified in \PerlmutterDS, is as follows:\vs

\bul States $(\L_+,0)$ remain dual to perturbative scalar excitations in the bulk. In the semiclassical limit the primaries have conformal dimension
\eqn\sua{h(\L_+,0) \approx -\L\cdot\hat{\rho} = -\half\sum_{j=1}^{N}j(N-j)d_j }
where $d_j$ are Dynkin labels. (See Appendix B.2 for further definitions and group theoretic details.) The lowest such primary, $({\small\yng(1)},0)$, has $h({\small\yng(1)},0)={1-N\over 2}$ and is dual to the bulk scalar with $m^2=-1+N^2$.

From the results of \PR, one can see that $({\small\yng(1)},0)$ exhibits generalized free-field behavior in the semiclassical limit, just as it does in the 't Hooft limit: in particular, a four-point function of two such operators and their conjugates factorizes into products of two-point functions, and presumably this factorization happens for all $n$-point functions of this field and its conjugate. This is the first sign that the bulk is weakly coupled in the semiclassical limit.
\vs

\bul States $(0,\L_-)$ are dual to classical solutions of the $\l=-N$ Vasiliev theory. Indeed, in the limit,
\eqn\sub{h(0,\L) \approx -{C_2(\L)\over N(N^2-1)}c+O(1)~.}
The linear scaling with $c$ is compatible with an interpretation in terms of classical solutions.
The corresponding Euclidean Chern-Simons connections represent
smooth, asymptotically AdS configurations at zero temperature with a
contractible spatial cycle and nonzero higher spin charge: these are
the so-called `conical defects' \CastroIW.\foot{The name derives from the
fact that for some subset of these smooth solutions, there exists a
gauge in which the spacetime metric is that of a conical surplus,
i.e. a conical deficit with negative opening angle. Despite the
fuzzy logic we continue to use the name in accordance with previous
literature.}  Imposing smoothness of the bulk connection determines its higher
spin charges to match those of a $(0,\L_-)$ state in the CFT.\vs

\bul States $(\L_+,\L_-)$ are dual to scalar excitations in conical
defect backgrounds.\vs

In what follows we will add to the evidence for this duality presented in \refs{\GaberdielKU,\PerlmutterDS}, by matching certain infinite classes of CFT four-point functions involving scalar and defect operators to dual holographic calculations in the $\l=-N$ Vasiliev theory.

\newsec{A primer:  conical defects and scalar fields  in Vasiliev gravity}
Our next order of business is to review the essential ingredients in the bulk calculation of scalar two-point functions in the background of a conical defect. We refer the reader to \refs{\ProkushkinBQ,\CastroIW,\KrausUF} for source material.
\subsec{Conical defects}

The pure higher spin sector of the Euclidean signature bulk theory
is described by a connection $A$, which is a 1-form taking values in
the Lie algebra sl(N,$\IC$).  We choose the sl(2) subalgebra under which
the defining N dimensional representation of sl(N) reduces to the
spin $(N-1)/2$ irreducible representation of sl(2), and call the
corresponding  sl(2)    generators  $L_{0}$, $L_{\pm 1}$.   Explicit
$N\times N$ matrix representations for these generators are given in
components as
\eqn\aaa{\eqalign{(L_0)_{jj} &= {N+1\over 2}-j\cr
(L_1)_{j+1,j} &= -\sqrt{(N-j)j}\cr
(L_{-1})_{j,j+1} &=\sqrt{(N-j)j}\cr}}
with $j=1,\ldots,N$, and with all other components vanishing.  In particular we note that $L_0^\dagger = L_0$ and $L_1^\dagger = -L_{-1}$.

Asymptotically AdS boundary conditions correspond to choosing the highest weight gauge \CampoleoniZQ\
\eqn\aa{\eqalign{A& = \left(e^\rho L_1 - \sum_{n=1}^{N-1} e^{-n\rho} Q_n (L_{-1})^n\right) dz+L_0 d\rho~.  }}
The corresponding metric will then behave asymptotically as
\eqn\ab{ ds^2 \approx d\rho^2 + e^{2\rho} dz d\zb~~,\quad \rho \rt \infty,}
and the higher spin fields die off sufficiently rapidly so as to define normalizable fluctuations near infinity.  The coefficients $Q_n$ are identified (up to proportionality constants) as the higher spin charges; for example, $Q_1$ is the holomorphic component of the  boundary stress tensor.

It is standard and convenient to strip off the $\rho $ dependence by defining $a$ via
\eqn\ac{ A = b^{-1} a b + b^{-1} db~,\quad   b= e^{\rho L_0}~.}
We have
\eqn\ad{a = \left( L_1 - \sum_{n=1}^{N-1}  Q_n (L_{-1})^n\right) dz~.  }
We will also make reference to the Hermitian conjugate connection, or more precisely $\abar$ defined as
\eqn\ae{ \abar = -a^\dagger~.}
Explicitly,
\eqn\af{ \abar =  \left( L_{-1} - \sum_{n=1}^{N-1} (-1)^{n+1} Q_n^* (L_{1})^n\right) d\zb~.  }

We now turn to the construction of smooth gauge field configurations.   Writing the complex boundary coordinate as $z= \phi + i\tau$, we will identify the angular coordinate as $\phi \cong \phi +2\pi$ so that the conformal boundary metric is a cylinder.   Consider the holonomy of the connection around the angular cycle,
\eqn\ag{  H = P e^{\oint A}~. }
Taking the path at fixed $\rho$ we can instead work with $\widehat{h}$ defined as $H=b^{-1} \widehat{h} b$ and given by
\eqn\ah{  \widehat{h} = P e^{\oint a}~. }
We say that the connection is smooth provided $\widehat{h}$ is an element of the center of SL(N,$\IC$).   In particular, such a holonomy acts trivially on the higher spin fields, since the connection transforms in the adjoint representation.  On the other hand, we should note that we will also be introducing matter fields which do not transform in the adjoint representation, so we will refer to such holonomies as smooth rather than trivial.

We will restrict to the case of constant $Q_n$, in which case we have
\eqn\ai{ \widehat{h} = e^\omega~,}
with
\eqn\aj{ \omega = 2\pi a_z=  2\pi \left( L_1 - \sum_{n=1}^{N-1}  Q_n (L_{-1})^n\right)~.}
The condition that $\widehat{h}$ be a central element then translates into conditions on the eigenvalues of $\omega$, which we write as
\eqn\ak{ {\rm eig}(\omega)= 2\pi i (n_1, n_2, \ldots , n_N)~.}
One has the freedom to choose the $n_i$ to form a strictly decreasing set. Combined with the requirement that \aj\ be diagonalizable, and the SL(N,$\IC$) tracelessness constraint, the allowed values of $n_i$ are conveniently parameterized in terms of non-negative integers $r_i$ which have a direct connection to Young tableau data. Taking $r_j$ to denote the number of boxes in the $j$'th row of the Young diagram,
\eqn\al{ n_j = r_j +{N+1\over 2} -{B\over N} -j~,}
with $B= \sum_j r_j$. Then $\widehat{h}=(-1)^{N+1} e^{-2\pi i B/N}
I$, which is a central element of SL(N,$\IC$), and the Young data are
those of the dual CFT state $(0,\L_-)$.

A choice of $\left\{r_j\right\}$ in principle specifies the values of the charges $\left\{Q_n\right\}$, although no explicit formula is known for the general case.  Happily, it will turn out that we will not need to use the explicit values of the charges.

\subsec{Scalar dynamics}

A propagating scalar field is described by the master field $C$, which obeys
\eqn\am{ dC+ A C - C \Ab =0.}
For the moment, we will take $A$ to be in the $N\times N$ defining matrix representation of sl(N,$\IC$), in which case $C$ is a general $N\times N$ matrix; we will discuss other representations later. The physical scalar field $\Phi$ is identified with the trace of the master field $C$,
\eqn\as{ \Phi = \Tr (C)~.}

  We solve  equation \am\ using the same technique employed in \KrausUF.  Since the connections are flat, we can write
\eqn\an{\eqalign{ A &= g^{-1} dg ~, \quad \Ab  = \gb^{-1} d\gb~.}}
We can take
\eqn\ao{\eqalign{ g& = e^{\Lambda_0} b~, \quad \gb  = e^{\Lamb_0}
b^{-1} }}
with 
\eqn\ap{\eqalign{ \Lambda_0 &= a_{z} z+a_{\zb}\zb ~, \quad \Lamb_0  = \abar_{z} z +  \abar_{\zb}
\zb~.}}
It is also useful to define
\eqn\aq{\eqalign{ \Lambda_\rho & = b^{-1} \Lambda_0 b=A_z z+A_{\zb}\zb~, \quad
\Lamb_\rho  = b \Lamb_0 b^{-1}= \Ab_{z} z+\Ab_{\zb}\zb~.}}
Gauge invariance then implies that solutions of \am\ are given by
\eqn\ar{\eqalign{ C& = g^{-1} c \gb \cr
&=e^{-\Lambda_\rho} b^{-1} c b^{-1} e^{\Lamb_\rho}~,}}
where $c$ is any constant (with respect to spacetime coordinates) matrix. The idea is that the gauge transformation $g$ sets $A=0$, and in this gauge the equation \am\ just implies constancy of $C$, and we call the solution $c$.   The physical $C$ is then obtained by  transforming back to physical gauge.

As noted above, any choice for the matrix $c$ leads to a solution of the scalar wave equation.  We are however interested in a particular solution, namely the one corresponding to the bulk-boundary propagator.   The correct choice for $c$ turns out to be a highest weight state. Namely, we demand
\eqn\at{\eqalign{
L_{-1}c& = c L_1 =0 \cr
 L_0 c &= cL_0 =-hc~. }}
Using the matrices in \aaa\ we find that $c_{ij} = \delta_{i,1}\delta_{j,1}$ and
\eqn\au{ h=-{N-1\over 2}~.}
As we will see, $h$ is equal to the holomorphic conformal dimension of the operator dual to $\Phi$.  Note that $h<0$, which is a reflection of the non-unitarity of the dual CFT.

With this choice of $c$ we readily compute
\eqn\av{ \Phi = e^{2h \rho}\Tr \left[ e^{-\Lambda_\rho} c e^{\Lamb_\rho}\right]~.}
This is the bulk-boundary propagator with the boundary point  at the origin, $z=\zb=0$.  Using the standard rules of AdS/CFT, the boundary two-point function $G(z,\zb)$ is read off from the behavior of the propagator near the boundary,
\eqn\aw{ \Phi = e^{-2h\rho} G(z,\zb) + \ldots~,\quad \rho \rt \infty~.}

A notable point regarding the above construction is that the computation of the correlator $G(z,\zb)$ is reduced to manipulations involving $N\times N$ matrices.  In particular, no differential equations need to be solved.

The simplest example is to take $Q_n=0$, which yields pure AdS in the coordinates \ab, and vanishing higher spin fields.  Since this corresponds to CFT on the plane, we will not periodically identify $z$ in this case. Recall that the bulk-boundary propagator for a field dual to a CFT operator of dimension $(h,h)$ is (up to normalization)

\eqn\ax{ G_{b\p}(\rho,z,\zb)= \left( e^{-\rho} \over e^{-2\rho}+z\zb \right)^{2h}~.}
On the other hand, evaluating \av\ with $\Lambda_\rho = e^\rho L_1 z$ and $\Lamb_\rho= e^{\rho} L_{-1}\zb$ we find
\eqn\ay{ \Phi =  \left( e^{-\rho}\over e^{-2\rho}+z\zb\right)^{-(N-1)}~.}
 This matches \ax\ with the value of $h$ given in \au.   The boundary two point function is read off as in \aw,
\eqn\az{ G(z,\zb) \propto  (z\zb)^{N-1}~.
}

\subsec{A shortcut to boundary two-point functions}
Although we can easily compute the full bulk-boundary propagators as above, one can compute the boundary correlator directly using the following insight.  According to \av-\aw\ this is
\eqn\ba{\eqalign{ G(z,\zb) &=\lim_{\rho \rt \infty} e^{-2(N-1)\rho}\Tr \left[ e^{-\Lambda_\rho} c e^{\Lamb_\rho}\right]\cr
& =\lim_{\rho \rt \infty} e^{-2(N-1)\rho}\sum_n \langle n|  e^{-\Lambda_\rho}|1\rangle \langle 1 | e^{\Lamb_\rho}|n\rangle  ~,}}
where we have passed to bra-ket notation.   Trading $\Lambda_\rho$
for $\Lambda_0$ using \aq, as well as
\eqn\bbb{ L_0 |n\rangle = {1\over 2}(N+1-2n)|n\rangle~,}
we find that only the $n=N$ term in \ba\ survives the $\rho \rt
\infty$ limit.  Using also that $\Lamb_0= - \Lambda_0^\dagger$, we
arrive at
\eqn\bc{ G(z,\zb) = \left| \langle N| e^{-\Lambda_0}|1\rangle \right|^2~.}
So the boundary correlator is read off from a single element of the matrix $e^{-\Lambda_0}$.

\newsec{Matching four-point functions}

We now show how to generalize the computation that led to \az\ to
the case where the background is an arbitrary smooth defect
solution.  The answer turns out to take a remarkably compact form.
This correlator should equal a CFT vacuum four-point function
involving two defect and two scalar operators, in the semiclassical
limit. We then perform this CFT computation; in the semiclassical limit, the
results will be seen to agree.

Recall the map laid out in Section 2 between
bulk scalar fields and conical defect geometries on the one hand,
and primary states of the CFT on the other. In particular, let us
define a shorthand for the $({\small\yng(1)},0)$ and
$(0,\L_-)$ primary operators --- corresponding to the
$m^2=-1+\l^2$ scalar field and defect geometries, respectively ---
as
\eqn\fpa{\phi \sim  ({\small\yng(1)},0)~,\quad D \sim (0,\L_-)}
with an obvious extension to their charge conjugates. The
representation $\L_-$ is specified by a Young diagram whose data is
encoded in the components $n_i$ introduced earlier.

The bulk scalar two-point function $G(z,\zb)$ in the defect background, which we now compute, is equivalent to a CFT vacuum four-point function: thinking of the defect background in terms of in and out states generated by $D$ acting on the vacuum, one has% 
\eqn\fpb{G(z,\zb) =  \lim_{z_1\rar\infty} |z_1|^{4h_D}\langle 0|D(z_1)\overline{\phi}(1)\phi(z)\overline{D}(0)|0\rangle}
where $h_D$ is the holomorphic conformal weight of the $D$ operator. The operation  $ \lim_{z_1\rar\infty} |z_1|^{4h_D}D(z_1)$  defines the operator $D'(z'=0)$, where $z'=1/z$.  Note that since $z$ is essentially the conformal cross ratio, conformal invariance places no constraints on the functional form of $G(z,\zb)$.

\subsec{Gravity}

We are interested in the boundary two-point function in the presence of the defect, so we need only compute the matrix element \bc\ for the conical defect background. This will correspond to a correlator on the cylinder; in this section we use $w$ to denote the cylinder coordinate, retaining $z$ for the coordinate on the plane.  Recalling \aj\ and \ak, the eigenvalues of $\Lambda_0$ are
\eqn\bd{ {\rm eig}(\Lambda_0) = i(n_1, n_2, \ldots , n_N)~.}
If the eigenvalues are all distinct then we can diagonalize,
\eqn\bee{ \Lambda_0 = U D U^{-1}~,}
which gives
\eqn\bfz{\langle N| e^{-\Lambda_0}|1\rangle = \sum_{j=1}^N U_{Nj}
U^{-1}_{j1} e^{-iw n_j}~.}

We compute the needed elements of $U$ following ideas in \CastroIW.  We first note that it is possible to find an upper triangular matrix $B$ obeying
\eqn\bg{ B^{-1} \Lambda_0 B   = \left( \matrix{ 0 & u_1 & u_2 &
\ldots & u_{N-1}&  u_{N-1} \cr -1 &0 & 0 & \ldots & 0& 0 \cr
 && \cdots &&&  \cr
 0 & 0 & 0& \ldots & -1 & 0  } \right) \equiv K~.  }
A count of the free parameters shows that there are precisely enough parameters in $B$ to zero out the appropriate entries of $K$.
For present purposes we will only need the following information about $B$:
\eqn\bh{ B_{NN}= (N-1)!B_{11}~.}
This relation is derived in  Appendix A.  The virtue of introducing $K$ is that it can be diagonalized by a Vandermonde matrix,
\eqn\bi{ K = VDV^{-1} }
where
\eqn\bj{ V = \left( \matrix{\lambda_1^{N-1} & \lambda_2^{N-1} & \ldots & \lambda_N^{N-1} \cr
&& \ldots & \cr
\lambda_1 & \lambda_2 & \ldots & \lambda_N \cr
1 & 1& \ldots & 1    } \right)~,  }
and $\lambda_j$ are the eigenvalues of $K$, and hence of $\Lambda$,
\eqn\bk{ \lambda_j = i n_j~.}

We have succeeded in writing $U=BV$, and since $B$ and $B^{-1}$ are upper triangular we have
\eqn\bl{\eqalign{ U_{Nj}&= B_{NN} V_{Nj} = B_{NN} \cr
U^{-1}_{j1}&= V^{-1}_{j1} B^{-1}_{11}~.}}
For a triangular matrix we have $B^{-1}_{11}=1/B_{11}$.  Also, the inverse of a Vandermonde matrix is well known  and obeys
\eqn\bm{ V^{-1}_{j1} = {(-1)^{N-1} \over \prod\limits_{l\neq j} (\lambda_l - \lambda_j)}~.   }
Putting these facts together we have,
\eqn\bn{ U_{Nj} U^{-1}_{j1} = {B_{NN} \over B_{11}} V^{-1}_{j1} = (N-1)! V^{-1}_{j1}=   {(-1)^{N-1}(N-1)!  \over \prod\limits_{l\neq j} (\lambda_l - \lambda_j)}}
and so
\eqn\boo{ \langle N | e^{-\Lambda_0} |1\rangle = \sum_{j=1}^N   {i^{N-1}(N-1)!  \over \prod\limits_{l\neq j} (n_l - n_j)} e^{-iw n_j }~.}
Our final result for the boundary correlator is then
\eqn\bp{ G(w,\wb) = \left|(N-1)! \sum_{j=1}^N {e^{-in_j w} \over \prod\limits_{l\neq j} (n_l - n_j)}\right|^2~. }

This result is on the cylinder. For eventual comparison to the CFT, we bring this to the plane, with coordinate $z$, using
\eqn\bpb{ G(z,\zb)= (z\zb)^{-h}G(w,\wb)|_{e^{-iw}=z} = (z\zb)^{N-1\over 2}G(w,\wb)|_{e^{-iw}=z}}
to give
\eqn\bpa{ G(z,\zb) = \left|(N-1)! ~z^{N-1\over 2}\sum_{j=1}^N {z^{n_j } \over \prod\limits_{l\neq j}(n_l - n_j)}\right|^2~. }
This is one of the main results of this paper. 

\vs
\noindent \bul {\it Example: defect in defining representation}\vs

The simplest nontrivial example corresponds to the choice
\eqn\bq{ r_1=1~,\quad r_2 = r_3 = \ldots = r_N =0~.}
Using
\eqn\brz{ \eqalign{  \prod_{l\neq 1}^N (n_l-n_1) & =    (-1)^{N-1} N! \cr
\prod_{l\neq j}^N (n_l-n_j)  & ={(-1)^{N+j}  (N-j)!j!\over j-1 } ~,\quad (j>1)    }}
and carrying out the sum over $j$ we find the following result on the cylinder
\eqn\bs{\eqalign{ G(w,\wb)= \left|{1\over N} e^{-{N^2 +N-2 \over
2N}iw}(e^{iw}-1)^{N-1}\big( (N-1)e^{iw}+1 \big)  \right|^2~,  }}
We can compare this to the corresponding CFT result computed in \PR\ (see their formula 3.10).
To compare, we here take $z=e^{iw}$, rather than the convention employed in \bpb.  This gives, after some rewriting
\eqn\bsa{ G(z,\zb)= \left| (z-1)^{N-1} z^{1\over N}  \left(1+{1-z \over Nz}\right) \right|^2~,}
which matches the result in {\PR}, noting that $\Delta_+ = -(N-1)$.

\subsec{CFT}

Our goal is to compute  the four-point functions on the right-hand side of \fpb\ from the CFT. We will use the Coulomb gas method to compute at arbitrary $(N,\l)$, and take the semiclassical limit at the end.

The specific cases $\L_-={\small\yng(1),\yng(1,1)}$ were computed in \PR,\foot{For instance, in \PR, they denote
$(0,{\small\yng(1)}) \sim \phi_-$ and $ (0,{\small\yng(1,1)})\sim
\phi_-^2$ with analogous expressions for operators
$({\small\yng(1)},0)$ and $({\small\yng(1,1)},0)$.} and this correlator was computed in some generality elsewhere, e.g. \refs{\FateevAB,\ChangYin}. We compute from first principles, with the main new insights being the dramatic simplification due to the choice of operators, and an even further simplification in the semiclassical limit. We
include a brief exposition of the Coulomb gas technique and the basic
technology required in Appendix B, along with details of the
following calculations. In what follows,
$\langle \cdot \rangle \equiv \langle 0| \cdot |0\rangle$.

\vs
\noindent \bul {{\it Defect in rectangular representation}}\vs

To warm up, we first treat the simpler case where the defect is in a
``rectangular'' representation, by which we mean a Young diagram
with $m$ boxes in each of $n$ columns. The SU(N) weight vector for
this representation  is $\L_- = n\o_m$, and we denote the defect
operator as $D_{n\times m}$.  This is
the simplest case from a computational point of view because the
weight vector $\L$ is proportional to a single fundamental weight of
SU(N).

As shown in the appendix, the procedure is straightforward given the simplicity of the representation, and the result for this correlation function, before taking any limit, is
\eqn\rda{\langle D_{n\times m}(\infty)\overline{\phi}(1)\phi(z)\overline{D}_{n\times m}(0)\rangle \propto \Big|(1-z)^{2\a_+^2\over N}z^{-{nm\over N}}{}_2F_1(\b,\gamma;\delta;z)\Big|^2}
where
\eqn\rdaa{\beta = 2\a_+^2 (N-m) + 1+m-N~, \quad \gamma = 2\a_+^2N+1-n-N~, \quad \delta = 2\a_+^2 + 1+m-N-n}
with
\eqn\rde{2\a_+^2 = {N+k+1\over N+k}~.}
Note the holomorphic factorization. We are not concerned with overall normalization here, as both this and the bulk results can be rescaled independently.

This matches the bulk result for general $(m,n)$ in the semiclassical limit $\a_+^2\rar 0$. First, we note that the arguments of the hypergometric function \rdaa\ are such that in the semiclassical limit it breaks up into a pair of finite sums: the correlator becomes
\eqn\rdf{\eqalign{&\lim_{k\rar -N-1} \langle D_{n\times m}(\infty)\overline{\phi}(1)\phi(z)\overline{D}_{n\times m}(0)\rangle\cr&\propto  \Big|z^{-{nm\over N}}{}_2F_1(1+m-N,1-n-N;1+m-N-n;z)\Big|^2\cr
&\propto \Bigg|z^{-{nm\over N}}\Bigg[\sum_{k=0}^{m-1}{\Gamma(1+n+k)\Gamma(1+k-m)\over \Gamma(1+N+n-m+k)}{z^{k+N+n-m}\over k!}\cr &\quad \quad \quad + \sum_{k=0}^{N-m-1}{\Gamma(1+m-N+k)\Gamma(1-N-n+k)\over \Gamma(1+m-N-n+k)}{z^k\over k!}\Bigg]\Bigg|^2}}
again up to overall constants which we have dropped.

Turning now to the bulk result, we bring it to the plane via the coordinate transformation $z=e^{-iw}$ and write the components $n_i$ explicitly in terms of the row data, $r_i$:
\eqn\rdg{G(z,\zb) \propto \Big|z^{-{nm\over N}}\sum_{j=1}^N {z^{r_j+N-j}\over \prod_{l\neq j}(l-j+r_j-r_l)}\Big|^2~.}
We split this into two sums: one over the first $m$ rows with $n$ boxes, and one over the remaining $N-m$ rows with zero boxes:
\eqn\rdh{G(z,\zb) \propto \Bigg|z^{-{nm\over N}}\left[\sum_{j=1}^{m}{z^{n+N-j}\over \prod_{l\neq j}(l-j+n-r_l)} + \sum_{j=m+1}^N{z^{N-j}\over \prod_{l\neq j}(l-j-r_l)}\right]\Bigg|^2~.}
Next, we shift the sums so that they start from zero; then,
for easier comparison to \rdf, write $j=m-1-k$ in the first sum, and $j=N-m-1-k$ in the second sum. \rdh\ becomes
\eqn\rdhb{G(z,\zb) \propto \Bigg|z^{-{nm\over N}}\left[\sum_{k=0}^{m-1}{z^{k+N+n-m}\over \prod_{l\neq m-k}(l+n-m-r_l+k)}+\sum_{k=0}^{N-m-1}{z^{k}\over \prod_{l\neq N-k}(l-r_l-N+k)}\right]\Bigg|^2}
Upon performing some Pochhammer gymnastics in the denominators, one finds precisely \rdf\ (up to overall constants).

For the simplest case in which the defect lives in the defining representation, the result is easily seen to match \bs.

\vs
\noindent \bul {\it Defect in general representation}\vs

We now turn to the case of an arbitrary defect representation
$D_{(0,\L_-)}$, with highest weight vector $\L_- =
\sum_id_i\o_i$. In Appendix B.3, we show that the
result before taking the semiclassical limit  reads
\eqn\axa{ \eqalign{ &\langle
D_{(0,\L_-)}(\infty)\overline{\phi}(1)\phi(z)\overline{D}_{(0,\L_-)}(0)\rangle
\propto\cr& \Big|C_{\L_-}
(1-z)^{{{2\alpha_+^2}\over{N}}}z^{v_{N-1}-{{B}\over{N}}-2\alpha_+^2}
{}_NF_{N-1}\left(
{{2\alpha_+^2,2\alpha_+^2\bf{1}-\bf{v}}\atop{\bf{1}-\bf{v}}}
\Big|{{1}\over{z}}\right) \Big|^2} , }
where
\eqn\axaa{ C_{\L_-} \equiv
\prod\limits_{k=1}^{N-1}\left({\Gamma(1-2\alpha_+^2)\Gamma(2\alpha_+^2-v_k)}\over{\Gamma(1-v_k)}\right)}
and $\bf{v}$ is an $(N-1)$-vector with components $v_k=\sum\limits_{j=1}^{k}\left(1+d_j-2\alpha_+^2\right)$.

Before taking the semiclassical limit it is important to address the
holomorphic factorization. The result is in principle expected to be
a sum of contributions from $N$ conformal blocks. If $|H_j(z)|^2$ is
the contribution of the $j^{th}$ conformal block, the
general expression for the correlator reads
\eqn\axa{ \eqalign{ \langle
D_{(0,\L_-)}(\infty)\overline{\phi}(1)\phi(z)\overline{D}_{(0,\L_-)}(0)\rangle
= & \sum\limits_{j=1}^{N} {\cal M}_{j j}|H_{j}(z)|^2 } , }
where ${\cal M}_{j j}$ is a constant factor fixed by demanding
monodromy invariance of the four-point function. Explicitly, this is
given, for our correlator, by \ChangYin
\eqn\axb{ \eqalign{ {\cal
M}_{j j} \propto & \prod\limits_{i=1,i\neq j}^{N}{{\Gamma\left(
\sum\limits_{k=j}^{i-1}(2\alpha_+^2-1-d_k) \right)\Gamma\left(
2\alpha_+^2- \sum\limits_{k=j}^{i-1}(2\alpha_+^2-1-d_k)
\right)}\over{\Gamma\left(         1-2\alpha_+^2+
\sum\limits_{k=j}^{i-1}(2\alpha_+^2-1-d_k) \right)\Gamma\left(    1-
\sum\limits_{k=j}^{i-1}(2\alpha_+^2-1-d_k) \right)}} } . }

Upon careful inspection one can argue that only the $N^{th}$ conformal block contributes. For $j<N$, the factor with $i=j+1$ will read
\eqn\axbb{
{{\Gamma\left(         2\alpha_+^2-1-d_j           \right)\Gamma\left(      1+d_k        \right)}\over{\Gamma\left(         -d_k  \right)\Gamma\left(    2- 2\alpha_+^2+d_k      \right)}} ~.
}
The second gamma function in the numerator will be an integer while the first gamma function in the denominator will diverge.  Consequently, ${\cal M}_{j j}=0$ for $j\neq N$.

We now take the limit $2\alpha_+^2\rar 0$ in \axa. Expanding the hypergeometric function in powers of $z$ the correlator becomes
\eqn\axc{\! \langle
D_{(0,\L_-)}(\infty)\overline{\phi}(1)\phi(z)\overline{D}_{(0,\L_-)}(0)\rangle
 \propto \Big|z^{v_{N-1}-{{B}\over{N}}}
\left(\prod\limits_{k=1}^{N-1}{{1}\over{v_k}}\right)
\sum_{i=0}^{\infty}\left[{z^{-i} (0)_i\over i!} \prod_{k=1}^{N-1}\left({{\left(-v_k\right)_i}\over{\left(1-v_k\right)_i}}\right) \right]
\Big|^2
}
Due to the $(0)_i$ Pochhammer symbol, the $i$'th term of the sum will vanish unless one of the Pochhammer symbols in the denominator vanishes too. This happens exactly $N$ times in the sum over $i$ at the points $i=0,v_k$. The $N$ contributions to the sum can then be written as
\eqn\axd{ \langle
D_{(0,\L_-)}(\infty)\overline{\phi}(1)\phi(z)\overline{D}_{(0,\L_-)}(0)\rangle
 \propto \Big|z^{-{{B}\over{N}}}
\sum_{j=1}^{N} {{z^{-(v_{j-1}-v_{N-1})}}\over{\prod\limits_{l\neq
j}\left(v_{j-1}-v_{l-1}\right)}} \Big|^2 . }
Going back to row data using $v_{i-1}-v_{N-1}=i-N-r_i$ and recalling the definition \al, the final result can be written
\eqn\axe{ \langle
D_{(0,\L_-)}(\infty)\overline{\phi}(1)\phi(z)\overline{D}_{(0,\L_-)}(0)\rangle
 \propto \Big|z^{N-1\over 2}
\sum_{j=1}^{N}
{{z^{n_j}}\over{\prod\limits_{l\neq j}\left(n_l-n_j\right)}}
\Big|^2 .
}
This equals the bulk result \bpa\ up to overall normalization.

\newsec{Scalars in general  representations}

Let us summarize the connection between CFT scalar operators and bulk master fields reviewed in Section 2.  Solutions of  the scalar  equation \am, where $A$ is a matrix in the defining representation of sl(N,$\IC$), map to CFT correlation functions involving a scalar operator in the coset representation $({\small \yng(1)},0)$.  In other words, the master field $C$ is dual to the CFT  primary $({\small \yng(1)},0)$ and its descendants.   This raises the obvious question: since the CFT contains more general scalar operators $(\Lambda_+,0 )$, what is the bulk dual of these operators?  Since such operators can be obtained from repeated OPEs of $({\small \yng(1)},0)$, we will refer to them as multi-trace operators, and then ask for the description of the corresponding multi-trace scalar in the bulk.

In ordinary AdS/CFT there is a standard prescription for handling such multi-trace operators \refs{\WittenUA,\BerkoozUG}.  Namely, one considers the original single trace scalar, but with modified boundary conditions.  However, in the framework implemented here, it is not so convenient to modify the boundary conditions, and so we propose a different approach.  There is a fairly obvious guess as to how to proceed:  since $({\small \yng(1)},0 )$ is dual to $C$ in the defining representation, it's only natural to surmise that $(\Lambda_+,0 )$ is dual to $C$ in the representation $\Lambda_+$.  By the latter, we mean that we consider \am\ with $A$ being a matrix in the representation $\Lambda_+$ of sl(N,$\IC$), and so $C$ is a  dim$(\Lambda_+) \times $ dim$(\Lambda_+)$ matrix.   In particular, we propose to employ the same method as before to compute bulk solutions,
\eqn\ava{ \Phi_{\Lambda_+} = e^{2h_{\Lambda_+} \rho}\Tr \left[ e^{-\Lambda_\rho} c e^{\Lamb_\rho}\right]~,}
and  boundary correlators,
\eqn\bt{G_{\Lambda_+}(z,\zb) = \lim_{\rho \rt \infty} e^{4h_{\Lambda_+} \rho} \Tr \left[ e^{-\Lambda_\rho} c e^{\Lamb_\rho}\right]~.}

We first check that the conformal dimension of the  CFT primary operator
$(\Lambda_+,0)$ is correctly reproduced. Let $d_i$ be the Dynkin labels of  $\Lambda_+$.   The CFT operator $(\Lambda_+,0)$ has conformal
dimension $(h_{\Lambda_+},h_{\Lambda_+})$ with
\eqn\bta{ h_{\Lambda_+}=   - \sum_{k=1}^N {k(N-k) \over 2}d_k~.}
On the bulk side, just as before,  $h_{\Lambda_+}$ is the $L_0$ eigenvalue of the highest weight $c$ matrix,
\eqn\ata{\eqalign{
T_{-n}c& = c T_n =0 \quad n>0~, \cr
 L_0 c &= cL_0 =-hc~. }}
Here $T_n$ denotes an sl(N) generator with weight $n$,  $[L_0, T_n]=-nT_n$.  Note that unlike for the defining representation we need that $c$ is annihilated by all the $T$ operators, since the conditions $L_{-1} c = cL_1=0$ have multiple solutions, as we discuss further below.

First consider a representation asym$_k$ which in Young tableau
language is a single column of $k$ boxes.  This has Dynkin labels
$d_k=1$ and $d_{m\neq k}=0$. Since this representation is the fully
anti-symmetric product of $k$ defining representations, the highest
weight state is
\eqn\bx{ |\hw\rangle \propto |1\rangle |2 \rangle \cdots |k\rangle \pm {\rm perms}~,}
where we are referring to our explicit matrix expressions for the defining representation.  In the defining representation the highest weight state is $|1\rangle$ with weight $h_{\small \yng(1)}=-(N-1)/2$. The column of $k$ boxes therefore has highest weight
\eqn\by{ h_{{\rm asym}_k}=-\sum_{l=0}^{k-1}\left( {N-1 \over 2}-l\right)= -{k(N-k)\over 2}~.}
For a representation with more than one column we just need to recall that the Young tableau symmetrizes the columns amongst each other and so the highest weight is additive. This leads to the following formula for the highest weight of a general representation
\eqn\bz{  h_{\L}= - \sum_{k=1}^N {k(N-k) \over 2}d_k~,}
in agreement with \bta.

It's also easy to show that the boundary correlator in Poincare AdS comes out correctly.   Following the same steps that led to \bc\ we  have
\eqn\bv{ G_{\Lambda_+}(z,\zb) = \left| \langle -\hw| e^{-\Lambda_0}
|\hw\rangle\right|^2~,}
with $\Lambda_0 =L_1 z$.   Since $L_1$ lowers the $L_0$ weight by
$1$ unit, only the $(L_1 z)^{-2h_{\Lambda_+}}$ term in the expansion of the
exponential contributes, and so
\eqn\bw{ G(z,\zb) \propto (z\zb)^{-2h_{\Lambda_+}}~,}
which is of course the expected result for an operator of conformal dimension $(h_{\Lambda_+},h_{\Lambda_+})$.

\subsec{Towers of scalar fields}
There is an important fact that needs to be emphasized: in general, a master field $C$ in a general representation $\Lambda_+$ describes a tower of scalar fields in AdS with various masses, and not just a single scalar field as was the case in the defining representation.  To appreciate this we should recall that fields in AdS are classified by irreducible representations of sl(2) $\times$ sl(2).  If we realize the sl(2) generators as the isometry generators of AdS, then the Klein-Gordon equation is simply the statement that the quadratic Casimir operators take a specified value.   The solutions of the wave equation transform in irreducible representations of  sl(2) $\times$ sl(2).    The master field $C$ is specified by a representation $\Lambda_+$ of sl(N); the key point is that $\Lambda_+$ decomposes into a sum of irreducible representation of sl(2), and each such sl(2) representation corresponds to a distinct field in AdS.     We recall that we are considering the principal embedding of sl(2) in sl(N), which is defined by the condition that the defining representation of sl(N) remains irreducible under the sl(2) subalgebra.    This fixes the branching of other representations, and thus specifies the spectrum of fields obtained for a given choice of $\Lambda_+$.   Each such sl(2) representation is dual to a CFT operator of dimension $(h_j, h_j)$, where $h_j=-j$ is the highest weight of the sl(2) spin $j$ representation.

As a concrete example, consider $\Lambda_+= {\small \yng(1,1)}$.  Labelling sl(2) representations by their spin $j$, with dimension $2j+1$, ${\small \yng(1,1)}$ decomposes as
\eqn\bwa{  {\small \yng(1,1)} \mapsto j = \left\{\matrix{(N-2) \oplus (N-4) + \ldots + (0) & (N~{\rm even})\cr  (N-2) \oplus (N-4) + \ldots + (1) & (N~{\rm odd})   }     \right. }
This also tells us the spectrum of highest sl(2) weights, namely $h = -(N-2),~ (N-4),~ \ldots$, and from this we read off the spectrum of masses via $m^2_j = 4h(h-1)=4j(j+1)$.    It is straightforward to verify that a field in a given sl(2) representation obeys the Klein-Gordon equation in AdS with the correct mass.    Let $|\psi_j\rangle$ be a state in a spin-$j$ representation.  The corresponding bulk solution is
\eqn\bwb{ \Phi_{\psi_j} = \langle \psi_j | b^{-1} e^{\Lamb_\rho} e^{-\Lambda_\rho} b^{-1} | \psi_j
\rangle ~.}
Just using properties of sl(2) we can verify
\eqn\bwc{ (\p_\rho^2 +2\p_\rho +4 e^{-2\rho} \p_z\p_{\zb} -m_j^2 )\Phi_{\psi_j}=0~,}
which is the relevant Klein-Gordon equation.

There is a parallel story on the CFT side, as the spectrum is fixed by sl(N) group theory. More precisely, the scalar fields in $C$ and their descendants furnish a non-unitary highest weight representation of sl(N), which is the large $c$ wedge subalgebra of the boundary $W_N$ conformal symmetry. To each bulk scalar field corresponds an sl(2) primary state of the CFT, and these are wedge descendants of the appropriate sl(N) primary state.

One can, for instance, package all of this information in the branching of the character of the sl(N) representation $\L$ into characters of sl(2) spin-$j$ representations. The sl(N) highest weight character is, for a representation $\L$ (see e.g. \GaberdielZW),
\eqn\gti{\chi_\L^{sl(N)} = q^{h_{\Lambda}}\prod_{i=2}^N\prod_{j=1}^{i-1}{(1-q^{\L_j-\L_i+i-j})\over (1-q^{i-j})}}
where $h_{\L}$ is the conformal weight \bz\ of the primary $(\L,0)$ in the semiclassical limit of the coset. The numerator of this expression captures the effect of the null states that render the representation finite-dimensional (at least for finite-sized Young tableaux). One can decompose the character as
\eqn\gtp{\chi_{\L}^{sl(N)}= \sum_{\lbrace j\rbrace} \chi_{j}^{sl(2)}}
where the $\lbrace j\rbrace$ are those determined by the branching rules. Each $ \chi_{j}^{sl(2)}$ contains the states dual to the subset of the fields in the bulk which obey the AdS Klein-Gordon equation with mass $m_j$. Executing this branching --- and explicitly constucting the physical and null states of a given representation --- is only a matter of algebra for low-lying representations.

We return to these ideas in Section 7, where we generalize them to the \hsl\ theory.

\newsec{Matching four-point functions II: Scalars in general representations}

We have just shown that in pure AdS, the two-point functions of scalar fields in general representations are those of the dual operators $(\L_+,0)$. This identification should apply in an arbitrary on-shell background, not only in AdS; we can use the same formula \bv,
\eqn\bva{ G_{\Lambda_+} = \left| \langle -\hw| e^{-\Lambda_0}
|\hw\rangle\right|^2~,}
where now $\Lambda_0$ refers to an arbitrary background. In particular, following the previous sections we would like to compute the two-point function for scalars in general representations in the background of a general conical defect.

It is easy to
see from the discussion of the previous subsection that the general result follows once we know the result for the scalar in the asym$_k$ representation, $G_{{\rm asym}_k}$.  Noting that $\Lambda_0$
acting on the tensor product representation is a sum of terms, one
for each factor, this implies that if we have $d_k$ columns of $k$ boxes we just
get $d_k$ factors of $G_{{\rm asym}_k}$.  The general result is
therefore
\eqn\caa{ G_{\Lambda_+} = \prod_{k=1}^N \big( G_{{\rm asym}_k} \big)^{d_k}~.}
Note that this holds in an arbitrary background, reference to which has been suppressed in the above.

Applied to the case of a conical defect background, the remaining work is therefore to compute $G_{{\rm asym}_k}$.  This appears to be quite tedious to obtain for general $k$, so in this next section we just consider the simplest nontrivial case, $k=2$, which will serve to illustrate the general approach.

\subsec{Gravity: $G_{{\rm asym}_2}$ with defect in general representation}

Using our explicit matrix representation for the sl(N) generators, the highest and lowest weight states in ${\small \yng(1,1)}$ are
\eqn\cb{\eqalign{ |\hw\rangle &= { |1\rangle |2 \rangle -  |2\rangle |1 \rangle \over \sqrt{2}} \cr
 |-\hw\rangle &= { |N\rangle |N-1 \rangle -  |N-1\rangle |N \rangle \over \sqrt{2}}~,}}
and so
\eqn\cc{ \langle -\hw | e^{-\Lambda_0} | \hw\rangle = \langle N  |
e^{-\Lambda_0} | 1\rangle \langle N-1  | e^{-\Lambda_0} | 2\rangle -
\langle N  | e^{-\Lambda_0} | 2\rangle \langle N-1  | e^{-\Lambda_0}
| 1\rangle~.}
Now, proceeding by the same route that arrived at \bfz, and recalling $U=BV$, the general matrix element takes  the form
\eqn\cd{ \langle m |e^{-\Lambda_0}|n\rangle = \sum_{jkl} B_{mj}
V_{jk} V^{-1}_{kl} B^{-1}_{ln}e^{-in_k w}~.}
In order to compute the matrix elements appearing in \cc\ we only need the following information about the upper triangular matrix $B$:
\eqn\ce{ B^{-1}_{ii}={1\over B_{ii}}~,\quad { B_{ii}\over B_{11}}= \sqrt{(N-1)!(i-1)!\over (N-i)!}~,\quad B_{i,i+1}=B^{-1}_{i,i+1}=0~.}
The relevant matrix elements are then
\eqn\cf{\eqalign{  \langle N  | e^{-\Lambda_0} | 1\rangle&=
(N-1)!\sum_k V_{Nk}V^{-1}_{k1} e^{-in_kw} \cr
 \langle N-1  | e^{-\Lambda_0} | 1\rangle&= \sqrt{(N-1)!(N-2)!}\sum_k V_{N-1,k}V^{-1}_{k1} e^{-in_kw} \cr
  \langle N  | e^{-\Lambda_0} | 2\rangle&=  \sqrt{(N-1)!(N-2)!}\sum_k V_{Nk}V^{-1}_{k2} e^{-in_kw} \cr
   \langle N-1  | e^{-\Lambda_0} | 2\rangle&= (N-2)!\sum_k V_{N-1,k}V^{-1}_{k2} e^{-in_kw}~.   }}
This gives
\eqn\cg{\eqalign{ \langle -\hw | e^{-\Lambda_0} | \hw\rangle &=
(N-1)!(N-2)!\sum_{jk}
\big(V_{Nj}V_{N-1,k}-V_{N-1,j}V_{Nk}\big)V^{-1}_{j1}V^{-1}_{k2}e^{-i(n_j+n_k)w}\cr
& = i(N-1)!(N-2)!\sum_{jk}
(n_k-n_j)V^{-1}_{j1}V^{-1}_{k2}e^{-i(n_j+n_k)w}~.  }}
Next we use
\eqn\ch{\eqalign{  V^{-1}_{j1} &= \prod_{k\neq j} {1 \over (\lambda_j - \lambda_k)} \cr
  V^{-1}_{j2} &= -  { \sum_{i}\lambda_i-\lambda_j \over \prod_{k\neq j} (\lambda_j - \lambda_k)} }}
to obtain
\eqn\ci{ \eqalign{\langle -\hw | e^{-\Lambda_0} | \hw\rangle
&=(-1)^{N-1} (N-1)!(N-2)!\sum_{jk} (n_k-n_j)\left(\sum_{i}n_i-n_k
\right)\cr
 &\quad\quad\quad  \times \prod_{m\neq j} {1 \over (n_j - n_m)}\prod_{n\neq k}{1\over  (n_k - n_n)}e^{-i(n_j+n_k)w} ~.}}
The term proportional to $\sum_i n_i$ is odd under $j\leftrightarrow k$ and so doesn't contribute.  The remaining terms can be written in a symmetric form as
\eqn\cj{ \eqalign{&\langle -\hw | e^{-\Lambda_0} | \hw\rangle =\cr &
(-1)^{N} (N-1)!(N-2)!\sum_{j<k} (n_k-n_j)^2\prod_{m\neq j} {1 \over
(n_j - n_m)} \prod_{n\neq k}{1\over  (n_k - n_n)}e^{-i(n_j+n_k)w}
~.}}
Using this result in \bva\ gives  our result for $G_{{\rm asym}_2}$.
\vs
\noindent \bul {\it Example:  $G_{{\rm asym}_2}$ with defect in defining representation}\vs

To obtain an expression that can be compared with our CFT computations, we now restrict to the case that the defect is in the defining representation,
\eqn\bqa{ r_1=1~,\quad r_2 = r_3 = \ldots = r_N =0~.}
We proceed to  evaluate the sums and products in \cj.

We  compute the $j=1$ and $j>1$  terms in \cj\ separately. The $j=1$ result is
\eqn\ck{\eqalign{\! -{e^{-{N^2+N-2\over N}iw}\over N^2(N-1)} \sum_{k=2}^N (-1)^k(k-1) k^2 { N! \over (N-k!)k!} e^{ikw} &= {e^{-{N^2-N-2\over N}iw}\over N}\left(Ne^{iw}-2\right)\left(1-e^{iw}\right)^{N-3} }}
The $j>1$ terms give
\eqn\cl{\eqalign{&{e^{-{N^2+N-2\over N}iw}\over N^2(N-1)} \sum_{j<k, j>1}^N (k-j)^2 {(-1)^j (j-1)N!\over (N-j)!j!} {(-1)^k (k-1)N!\over (N-k)!k!}e^{i(j+k)w}\cr
&  =-{e^{-{N^2-N-2\over N}iw}\over N}\big(1-e^{iw}\big)^{N-4}\Big( \big((N-2)e^{iw}+2\big)\big(1-e^{iw}\big)^N +(Ne^{iw}-2)(1-e^{iw})   \Big)~.   }}
Adding the results gives
\eqn\cla{\langle  -\hw | e^{-\Lambda_0} | \hw\rangle =
{2(-1)^{N-1}\over N} e^{-{N^2-N-2\over N}iw}
\left(1-e^{iw}\right)^{2(N-2)}\left(1+{N-2\over 2}e^{iw}\right)~.}

This yields the correlator on the cylinder according to $ G(w,\wb)= |\langle  -\hw | e^{-\Lambda_0} | \hw\rangle|^2$.
Comparison to the CFT is easiest if we map  this to the plane, $z=e^{-iw}$,
\eqn\cqa{ G(z,\zb) = \left|{N-2\over N}  z^{-{2\over N}} (1-z)^{2(N-2)}\left(1+{2\over N-2}z\right)\right|^2~.}
We now show that \cqa\ matches the corresponding CFT result.

\subsec{CFT}
We turn to the computation of the four-point function
\eqn\nfb{\langle D(\infty)\overline{\phi}(1)\phi(z) \overline{D}(0)\rangle}
where now \eqn\sta{\phi \sim(\L_+,0) ~, ~~ D \sim(0,{\small\yng(1)})~.}
Specifically, we focus on the case where $\L_+$ is a rectangular
representation, with weight vector $\L_+ = n\o_m$. The bulk
computation of the previous subsection is for the representtion
$n=1,m=2$, but we can be more general here. This is similar to the
calculation in subsection 4.2. Call the scalar field
$\phi_{m\times n}$. As we show in Appendix B.4, the correlator
\nfb\ is found to be
\eqn\stb{\langle D(\infty)\overline{\phi}_{n\times m}(1)\phi_{n\times m}(z) \overline{D}(0)\rangle \propto \Big|(1-z)^{-\Delta_{n\times m}}z^{-{nm\over N}}{}_2F_1(\b,\gamma;\delta;z)\Big|^2}
where
\eqn\stc{\b = -n~, \quad \gamma = -2\a_-^2m+m~, \quad \delta = 2\a_-^2(N-m) + 1+m-N-n}
with
\eqn\std{2\a_-^2 = (2\a_+^2)^{-1} = {N+k\over N+k+1}}
$\Delta_{n\times m}$ is the conformal dimension of the scalar representation,
\eqn\stda{\Delta_{n\times m} = -nm(N-m) + 2\a_+^2\left({nm(N-m)(n+2N)\over N}\right)~.}
Again we note that the result is holomorphically factorized.

In the semiclassical limit, $\a_-^2$ blows up, and \stb\  becomes
\eqn\nft{\eqalign{&\lim_{k\rar -N-1}\langle D(\infty)\overline{\phi}_{n\times m}(1)\phi_{n\times m}(z)\overline{D}(0)\rangle=\cr&\lim_{\a_- \rar\infty}\Big|(1-z)^{nm(N-m)}z^{-{nm\over N}}{}_2F_1(-n,-2\a_-^2m;2\a_-^2(N-m);z)\Big|^2~.}}
Notice that the series will terminate at $O(z^n)$; furthermore, the Pochhammer symbols in the series expansion degenerate into powers of their arguments:
\eqn\nfo{\lim_{\a_- \rar\infty}{(-2\a_-^2m))_p\over (2\a_-^2(N-m))_p} = {(-2\a_-^2m)^p\over (2\a_-^2(N-m))^p} = \left({-m\over N-m}\right)^p}
for integer $p$. Then
\eqn\nfp{\eqalign{&\lim_{k\rar -N-1}\langle D(\infty)\overline{\phi}_{n\times m}(1)\phi_{n\times m}(z)\overline{D}(0)\rangle\cr&= \Big|(1-z)^{nm(N-m)}z^{-{nm\over N}}\sum_{p=0}^{\infty} {(-n)_p\over p!}\left({-zm\over N-m}\right)^p\Big|^2\cr
&= \Big|(1-z)^{m(N-m)}z^{-{m\over N}}\left(1+{m\over N-m}z\right)\Big|^{2n}~.}}
This has the expected factorized form, symmetrized among the $n$ columns, reduces to the correct result for the defining representation in the $m=n=1$ limit, and matches the bulk calculation of the ${\small\yng(1,1)}$ representation  \cqa\  at $n=1,m=2$.

The result \nfp\ combined with group theory arguments given around
equation \bx\  suggests that the correlator for an arbitrary
scalar representation $\phi\sim(\L_+,0)$, with $\L_+ = \sum
d_j\o_j$, in the defect background $D\sim({\small\yng(1)},0)$
should be, in the semiclassical limit, \eqn\nfw{\langle
D(\infty)\overline{\phi}_{(\L_+,0)}(1)\phi_{(\L_+,0)}(z)\overline{D}(0)\rangle
\propto \Bigg|\prod_{j=1}^{N-1}\left[(1-z)^{j(N-j)}z^{-{j\over
N}}\left(1+{j\over N-j}z\right)\right]^{d_j}\Bigg|^2~.} However, we
have not actually computed this.

\newsec{Scalar fields and primaries in \hsl}

The identification of $(\L_+,0)$ CFT representations with the bulk
master field $C$ living in the $\L_+$ representation of SL(N)
extends to the case of \hsl, which we now explore in both CFT and
gravity. The main message is as follows: taking $C$ to live in any
representation but the defining one introduces an infinite tower of
scalar fields, and this matches what one sees from CFT. In checking
this conjecture, we will develop new tools for computing correlation
functions of scalar fields in the \hsl\ theory for generic $\l$, and
hence pertinent to the 't Hooft limit of the $W_N$ minimal models.

First, the CFT side. Let us recall some results of \GaberdielZW. Consider the characters for some low-lying representations of \hsl:
\eqn\gtx{\eqalign{\chi_{{\small\yng(1)}}^{{\rm hs}[\l]} &=q^{h}{1\over 1-q}\cr
\chi_{{\small\yng(2)}}^{{\rm hs}[\l]} &=q^{2h}{1\over (1-q)(1-q^2)}\cr
\chi_{{\small\yng(1,1)}}^{{\rm hs}[\l]} &=q^{2h+1}{1\over (1-q)(1-q^2)}\cr}}
where $h={1+\l\over 2}$.
The \hsl\ representations are infinite-dimensional, and we can think of \gtx\ as U($\infty$) characters, that is, the large $N$ limit of U(N) characters. Compared to the SL(N) character in \gti, the character of a U(N) representation lacks only an overall factor of $q^{-{BN\over 2}}$. The authors of \GaberdielZW\ use notation
\eqn\gtxb{P_{\L}(q) = \chi_{\L}^{U(\infty)}(z_i)~, ~~ z_i = q^{i-\half}}
to denote the large $N$ limit of the U(N) character, and
\eqn\gtxc{P_{\L}^{\pm}(q) = q^{{\pm\l B\over 2}}P_{\L}(q)~.}
With this notation, the \hsl\ character for an arbitrary representation is just
\eqn\gtxd{\chi_{\L}^{{\rm hs}[\l]} = P^{\pm}_{\L}(q)}
(cf. eqs. (3.12) and (6.10) of \GaberdielZW). The central point is that in the large $N$ limit, terms $q^N\rar 0$, which renders representations infinite-dimensional by removing numerator factors. Notice that $P_{\L}^{\pm}(q)$ is not quite the continuation of the SL(N) character to $N=\pm\l$, but it is close, and indeed intuitive. One starts with a state of highest weight and acts with an infinite number of lowering operators, which is exactly what one gets upon taking the large $N$ limit of the SL(N) characters.

One can apply this logic to derive the following expression as the character of an irreducible \hsl\ representation with highest weight $h_n$:
\eqn\gtxg{\chi_{h_n}^{{\rm hs}[\l]} = {q^{h_n}\over 1-q}}
where $h_n$ is some polynomial in $\l$.

In accord with earlier remarks, this immediately implies that in the bulk \hsl\ theory, the master field $C$ living in an arbitrary representation $\L$ contains an infinite tower of scalar fields each transforming under a non-half-integer spin representation of sl(2).

\subsec{Computing \hsl\ scalar propagators for generic $\l$}
To check this from the bulk, we want to compute two-point vacuum correlation functions of the $(\L_+,0)$ primaries from the \hsl\ Vasiliev theory, just as we did for sl(N) in Section 5. This requires an extension of the work \KrausUF\ which we detail now.

Let us treat the familiar case of $\L_+={\small\yng(1)}$ first, which we know should be described by only a single scalar field in the bulk with $m^2=-1+\l^2$ and bulk-boundary propagator
\eqn\buba{G_{b\p} = \left({ e^{-\rho}\over e^{-2\rho}+z\zb }\right)^{1+\l}~.}
 In fact, even to compute the vacuum correlator of this field from the bulk, we must improve upon the treatment in \KrausUF. There, technology was developed to efficiently compute these propagators at the special point $\l=1/2$, where there is a realization of the \hsh\ algebra in terms of monomials of harmonic oscillator variables which star-multiply via the Moyal product. While the conceptual framework is identical for generic $\l$, away from this point it was not well understood how to compute arbitrary traces of star products of exponential functions.

To be concrete, the scalar bulk-boundary propagator in a generic background $(A,\Ab)$ for generic $\l$ can be written as
\eqn\bua{\Phi = e^{(1+\l)\rho}\Tr[e^{- {\Lambda}_{\rho}}\star c \star e^{ \overline{\Lambda}_{\rho}}]}
where $\L_{\rho}$ was defined in \aq. The $A=0$ gauge master field $c$ is a highest weight state of \hsl\ (spanned by generators $\lbrace V^s_m\rbrace$), given explicitly as the infinite sum over zero modes
\eqn\bub{c = \sum_{t=1}^{\infty} c_t V^t_0~, ~~{\rm where}~~ c_t = {4^{t-1}(2t-1)\over (t-1)!}{\Gamma(\half)\over\Gamma({3\over 2}-t)}{\Gamma(\l+1-t)\over\Gamma(\l)}~.}
The star operation is defined here as the ``lone star product'' \PopeSR,
\eqn\buc{V^s_m \star V^t_n \equiv \half \sum_{u=1,2,3,...}^{s+t-|s-t|-1}g^{st}_u(m,n;\l)V^{s+t-u}_{m+n}}
where $g^{st}_u$ are structure constants polynomial in $\l$. By ``$\Tr$'' we mean the isolation of the $V^1_0$ component which acts as the identity. For basic material on \hsl, see e.g. \refs{\GaberdielWB, \KrausDS}.

In Poincar\'e AdS, say, the connection and hence $(\L_{\rho},\overline{\L}_{\rho})$ are simple, and the propagator is
\eqn\buaa{\Phi = e^{(1+\l)\rho}\Tr[e^{-ze^{\rho}V^2_1}\star c \star e^{\zb e^{\rho}V^2_{-1}}]~.}
Even in this case one must contend with infinite sums and an infinite number of nonzero traces to find $\Phi$, which is a challenge to compute. In addition, this construction does not easily generalize to larger representations $\L_+$ of \hsl. To see why, we note that if we wish to construct $C$ in the representation $\L_+$ by passage from $A=0$ gauge, we need to first find the field $c$ in that same representation. But as derived in \KrausUF, \bub\ is the unique highest weight state of \hsl\ subject to the lone star multiplication. Indeed, the lone star product is the analog of fundamental matrix multiplication of SL(N): a star product of two generators is written in \buc\ as a linear combination of all \hsl\ generators and the identity. So to understand larger representations of \hsl\ in this generator formalism, we must understand the other star multiplication rules that give rise to the \hsl\ star-commutation relations. This appears not to be a simple task.

However, there exists an infinite-dimensional matrix representation of \hsl\ which greatly simplifies matters in the present setting. First, recall an infinite-dimensional representation of sl(2),
\eqn\bud{\eqalign{(V^2_0)_{kk} &= {-\l+1\over 2}-k\cr
(V^2_1)_{k+1,k} &= -\sqrt{(-\l-k)k}\cr
(V^2_{-1})_{k,k+1} &=\sqrt{(-\l-k)k}\cr}}
where $k=1,\ldots,\infty$. One can think of this as a representation of sl(2) for arbitrary spin. From this we construct the full \hsl\ algebra via the enveloping algebra construction \PopeSR.
To recover sl(N), one sets $\l=-N$ and truncates the matrices to the upper left $N\times N$ block: the relation to the fundamental representation of sl(N) is thus manifest.

To evaluate the scalar propagator we need to understand what $c$ looks like in matrix form.
While it is completely unobvious from \bub, the \hsl\ highest weight state $c$ is given by the matrix

\eqn\bue{c = {\rm diag}(1,0,0,\ldots)~.}
This is clearly a \hsl\ highest weight state: in the sl(2) sector we have
\eqn\buf{\eqalign{V^2_0  c&=-\left({1+\l\over 2}\right)c=c  V^2_0\cr
V^2_{-1}  c &= c  V^2_1 =0}}
and the analogous equations are obeyed for all higher spin generators; this follows simply from the enveloping algebra construction.

Plugging \bue\ into \bua, the scalar propagator in a generic on-shell background simplifies to
\eqn\bug{\Phi = e^{(1+\l)\rho}[e^{ \overline{\Lambda}_{\rho}}  e^{- {\Lambda}_{\rho}}]_{11}~.}
Just as in the sl(N) case, we need only to compute a single matrix element. This observation avoids the need to directly take a trace on the space of infinite-dimensional matrices.\foot{There is a natural trace operation defined by \KhesinEY\ but it comes with certain subtleties, which we can happily circumnavigate here.}

In Poincar\'e AdS we compute \buaa\ using \bug. Expanding the exponentials and using, from \bud,
\eqn\buga{[(V^2_{-1})^p(V^2_1)^q]_{11}   = \delta_{p,q}{q!\Gamma(q+\l+1)\over \Gamma(\l+1)}~,}
we easily recover the correct result \buba.

One can also straightforwardly compute the propagator for generic $\l$ in the chiral deformation background of \KrausUF, the simplest higher spin deformation of Poincar\'e AdS. The result correctly reduces to that of \KrausUF\ at $\l=1/2$.
\vs

The previous technique can be generalized to deal with fields with generic sl(2) spin, whereas the sl(2) matrix representation \bud\ has highest weight $h=-(V^2_0)_{11} = {1+\l\over 2}$.
The following matrices obey the canonical sl(2) commutation relations and reduce to the defining representation \bud\ when $h={1+\l\over 2}$:
\eqn\buh{\eqalign{(V^2_0)_{kk} &=  -h-k+1\cr
(V^2_1)_{k+1,k} &= -\sqrt{(-2h-k+1)k}\cr
(V^2_{-1})_{k,k+1} &= \sqrt{(-2h-k+1)k}~.\cr}}
One can think of these as the continuation of the generic spin-$j(=-h)$ sl(2) representation to non-integer spin, via $N\rar-\l$. The master field $c$ of an sl(2) highest weight state with weight $h$ is again given by \bue.

In analogy with the preceding subsection, we can use these matrices to describe bulk scalar dynamics when $C$ lives in a general representation $\L_+$. For each of the infinite number of scalar fields living in the master field $C$, parameterized by some sl(2) weight $h_n$ as in \gtxg, we can compute the AdS scalar propagator using the representation \buh\ with $h=h_n$. The derivation exactly follows the case of the defining representation, except now one has
\eqn\bui{[(V^2_{-1})^p(V^2_1)^q]_{11}   = \delta_{p,q}{q!\Gamma(q+2h_n)\over \Gamma(2h_n)}}
with which we calcuate
\eqn\buj{\eqalign{\Phi &= \left({ e^{-\rho}\over e^{-2\rho}+z\zb}\right)^{2h_n}}}
which is correct. Thus, we have computed vacuum two-point functions of $(\L_+,0)$ states in the 't Hooft limit of the CFT, where $\l<1$, from the gravity side.

A perhaps more direct way to compute the propagator, especially for \hsl\ highest weight states, is to build higher representations of \hsl\ by taking tensor products of the defining representation as we did in \cb\ for sl(N). This reduces the computation of all matrix elements to those of the defining representation. For low-lying representations this is tractable, and one arrives the  correct results.

\newsec{Discussion}

We have presented new evidence for a duality between the semiclassical limit of the $W_N$ minimal models and the Vasiliev theory at $\l=-N$, beyond considerations of symmetry. Aside from lending support to this conjecture, our four-point function calculations are exciting from the perspective of the stronger form of the AdS/CFT correspondence. When we match quantities on both sides, there is no interpolation in the coupling, nor reliance on non-renormalization theorems: the coupling is the same on both sides, and the quantities literally match. It is widely believed that Vasiliev's theories of higher spin gravity are toy models for string theory at high energies. To the extent that this is true, our calculation is the moral equivalent of reproducing correlation functions in free ${\cal{N}}=4$ super Yang-Mills from a bulk computation of classical string theory in a strongly curved background. Obviously, it would be very desirable to investigate to what extent the 3D Vasiliev theory can actually be recovered from, or deformed to match, a limit of string theory. There are recent indications from holography that such a connection exists for a supersymmetric version of the 4D Vasiliev theory \ChangKT.

Both the 't Hooft and semiclassical limits of the CFT are interesting for their own reasons. The 't Hooft limit  incorporates an infinite tower of higher spin fields while maintaining unitarity, but the resulting spectrum contains the enigmatic light states.  The semiclassical limit reveals a spectrum that maps to states of a classical bulk theory that we understand fairly well, is more computationally tractable, and consequently holds promise that we may be able to compute many interesting quantities exactly on both sides.  However, the theory is non-unitary, raising questions about how to define the theory at the quantum level.

At first glance, the 't Hooft limit seems more conventional than the semiclassical limit with regard to the quantum loop expansion in the bulk. In particular, in the 't Hooft limit it is $1/N$ that controls the expansion, similar to what one has in more familiar examples of AdS/CFT.   On the other hand, in the semiclassical limit the rank $N$  is held fixed, and it is by tuning $k$ that we reach the classical limit in the bulk. This seems quite different.\foot{We thank Rajesh Gopakumar for his thoughts on this topic.} However, the triality symmetry of the quantum $W_{\infty}[\l]$ algebra suggests that the coset \ca\ possesses a generalized level-rank duality
\eqn\ccb{ {SU(N)_k \oplus SU(N)_1 \over SU(N)_{k+1} }\cong
 {SU(M)_{\ell} \oplus SU(M)_1 \over SU(M)_{\ell+1} }}
where $M$ is defined as in \cg, and $\ell = M/N-M$. In the semiclassical limit, $M\rar-\infty$, $\ell\rar+\infty$ and $c\sim -M\sim 1/G_N$. This, then, is quite similar to the 't Hooft limit in the above sense and seems to comport with the usual holographic paradigm, with the caveat that the rank $M$ is becoming negative rather than positive; the latter might reflect the non-unitarity of the limit.

There are various avenues one can pursue using the results herein. For instance, the CFT correlators we compute are exact in $N$ and $k$, and can be expanded order by order in $1/c$, providing a wealth of predictions for quantum corrections in the bulk.  It would of course be very interesting to learn how to reproduce these quantum effects via bulk perturbation theory.

 With regard to our proposal to introduce master fields in higher representations of \hsl, we must point out that so far we have only checked this proposal at the level of free scalar fields in the bulk.  A general open problem is whether it is possible to formulate the interacting scalar theory such that \hsl\ symmetry is manifest.   Given such a formulation, we could then ask about how master fields in higher representations fit into this framework. It is also interesting to ask whether a version of the semiclassical limit of the CFT holds for arbitrary $\lambda$; if so, then we could to extend the bulk/CFT agreement found here at $\lambda=-N$ to arbitrary $\lambda$.

Finally, we note that the tools of Section 7 for computing scalar correlation functions in the \hsl\ theory for generic $\l$ --- and for introducing master fields in higher representations of \hsl\ --- can be used to study scalar dynamics in the bulk in many contexts. For instance, it would be nice to generalize the calculations of scalar correlators in the \hsl\ higher spin black hole background \KrausDS\ to the case of generic $\l$ and/or to scalar fields in higher representations. In \KrausUF, the explicit correlator was only computed at $\l=1/2$ for the basic scalar field in the  defining representation. We can quickly note that tools developed here are sufficient, following Section 3 of \KrausUF, to show that the bulk-boundary propagator of a scalar field living in any representation will be thermally periodic in the higher spin black hole background.

\vskip .3in

\noindent
{ \bf Acknowledgments}

\vskip .3cm

We wish to thank Andrea Campoleoni, Constantin Candu, Slava Didenko, Matthias Gaberdiel, Rajesh Gopakumar, Shiraz Minwalla, Tomas Prochazka, Joris Raeymaekers, Mukund Rangamani and Evgeny Skvortsov for helpful discussions. E.P. wishes to thank the ETH, Zurich and the Solvay Workshop on Higher Spin Gauge Theories in Brussels for hospitality during the course of this work.  E.P. has received funding from the European Research Council under the European Union's Seventh Framework Programme (FP7/2007-2013), ERC Grant agreement STG 279943, “Strongly Coupled Systems”.  P.K. is supported in part by NSF grant PHY-07-57702. E.H. acknowledges support from  `Fundaci\'on La Caixa'.

\appendix{A}{Properties of $B$}

Here we derive some properties of the matrix $B$ that we referred to in Sections 4 and 6.
Recall that $B$ is defined to be an upper triangular matrix obeying
\eqn\da{ B^{-1} \Lambda B   = \left( \matrix{ 0 & u_1 & u_2 & \ldots & u_{N-1}&  u_{N-1} \cr -1 &0 & 0 & \ldots & 0& 0 \cr
 && \cdots &&&  \cr
 0 & 0 & 0& \ldots & -1 & 0  } \right) \equiv K~,  }
with
\eqn\db{
\Lambda =\left(  L_1 - \sum_{n=1}^{N-1}  Q_n (L_{-1})^n\right)z~.}

It's easy to show that the upper triangular property of $B$ implies the same for  $B^{-1}$, as well as
\eqn\dc{ B^{-1}_{ii} = {1\over B_{ii}}~,\quad B^{-1}_{i,i+1} =-{B_{i,i+1}\over B_{ii} B_{i+1,i+1}}~.}

The diagonal elements of $B$ can be determined from the condition $(B^{-1}\Lambda B)_{j+1,j}=-1$.  Since $L_{-1}$ is  nonzero only above the diagonal it follows that only $L_1$ contributes, and so we have
\eqn\dca{ -1=(B^{-1}L_1 B)_{j+1,j}= B^{-1}_{j+1,j+1}(L_1)_{j+1,i}B_{jj} = -\sqrt{j(N-j)} B^{-1}_{j+1,j+1}B_{jj}~.}
Using \dc\ leads to a recursion relation whose solution is
\eqn\dcb{B_{jj} = {\sqrt{(N-1)!(j-1)!\over (N-j)!} } B_{1,1}~. }

Similarly, the equation $(B^{-1}\Lambda B)_{jj}=0$ leads to
\eqn\hiaa{ 0 = B^{-1}_{ii}(L_1)_{i,i-1} B_{i-1,i}+B^{-1}_{i,i+1} (L_1)_{i+1,i} B_{ii}~, }
except that for $i=1$ the first term is absent.  The $i=1$ equation, together with \dc\ yields
\eqn\dcc{ B^{-1}_{12}=B_{12}=0~.}
But then \hiaa\ and \dc\ are easily seen to imply
\eqn\dcd{ B^{-1}_{j,j+1}=B_{j,j+1}=0~.}

\appendix{B}{Coulomb gas calculations}
We provide a brief description of the Coulomb gas method for computing correlation functions in minimal models, followed by some group theoretic data on SU(N) and some details of calculations in the main text. Our notation and methodology are guided by \refs{\PR, \DiFrancescoNK} which also contain nice pedagogical introductions.

\subsec{A brief review of the Coulomb gas method in $W_N$ minimal models}
The basic objects in the Coulomb gas method are vertex operators made from $N-1$ free bosons,
\eqn\cgl{V_{\bb} = : e^{i\bb\cdot \phi}:}
where $\bb$ is an $N-1$ vector. These bosons exist in the presence of a background charge which ensures that the central charge is that of the $W_N$ minimal model series (cf. Section 2). To each CFT operator corresponds a vertex operator of the same conformal dimension; this fixes $\bb$ via
\eqn\cgm{h_b = \bb\cdot (\bb-2\a_0\hat{\rho})}
where $\hat{\rho}$ is the Weyl vector of SU(N), and $\a_0$ is defined as
\eqn\cgla{\a_0= \sqrt{1\over 2(N+k)(N+k+1)}~.}
For a general $W_N$ primary $(\L_+,\L_-)$, the map $(\L_+,\L_-) \mapsto V_{\bb_i}$ is as follows:
\eqn\cgu{(\L_+,\L_-) \quad \mapsto \quad V_{\bb}~, ~~ \bb =
-\a_+\L_+-\a_-\L_-}
and $\L_{\pm}$ are the SU(N) weight vectors,
\eqn\cgva{\L_{\pm} = \sum_{i=1}^{N-1}d_{\pm,i}\o_i}
with $d_i$ the Dynkin labels and $\o_i$ the fundamental weights. One can confirm that this assignment gives the correct conformal dimension of the coset operators.

To compute a CFT four-point function, one maps the calculation to a correlation function of vertex operators in the free boson theory. To satisfy charge neutrality of the correlator in the presence of background charge, one must also insert a certain set of screening charges in the Coulomb gas correlator:
\eqn\cgn{Q = \oint dz V_{\tilde{\bb}}(z)~.}
These are nonlocal objects with nonzero charge, but $\tilde{\bb}$ is chosen such that their conformal dimension equals zero, and hence their insertion in a correlation function does not spoil conformal invariance. In particular, the constraint $h_Q=0$ forces
\eqn\cgo{\tilde{\bb} = \a_{\pm}\be_i}
where $\be_i$ is one of the simple roots of SU(N), and
\eqn\cgp{\a_{\pm} = \half(\a_0\pm\sqrt{\a_0^2+2})~.}
So there are $2(N-1)$ possible screening charges $Q_i^{\pm}$. For the purposes of taking the semiclassical limit, it is convenient to define $p=N+k$, in which case
\eqn\cgq{\eqalign{\a_0= \sqrt{1\over 2p(p+1)}~, ~~ \a_+ = {p+1\over \sqrt{2p(p+1)}}~, ~~ \a_- = -{p\over \sqrt{2p(p+1)}}}~.}
In the semiclassical limit, $p\rar -1$. Note that $\a_+\a_-=-\half$.

The determination of the screening charges is fixed by the charge conservation condition among all $\lbrace \bb\rbrace$ within a Coulomb gas correlator, which is
\eqn\cgq{\sum \bb = 2\a_0\hat{\rho}~.}
A crucial point is that $V_{\bb}$ and $V_{2\a_o\hat{\rho}-\bb^*}$ can be exchanged for one another in a Coulomb gas correlator, where $\bb^*$ is the conjugate representation of $\bb$. This can simplify the satisfaction of \cgq.

Thus, computing correlators via the Coulomb gas is essentially a four-step process: map the CFT operators to the appropriate Coulomb gas vertex operators (cf. \cgu); insert the correct number of screening charges to satisfy charge conservation (cf. \cgn, \cgq); evaluate the integrand using free-field contractions,
\eqn\chii{\langle V_{\bb_1}(z_1)\ldots V_{\bb_n}(z_n)\rangle= \prod_{i<j}|z_{ij}|^{4\bb_i\cdot \bb_j}~;}
and then integrate. The end result is the map
\eqn\cgs{\langle \Oc_1\Oc_2\Oc_3\Oc_4\rangle=\langle V_{\bb_1}V_{\bb_2}V_{\bb_3}V_{\bb_4}\prod_{i=1}^{N-1}(Q^+_i)^{m_i}(Q^-_i)^{n_i}\rangle}
where we must choose the $(m_i,n_i)$ such that the charge conservation condition
\eqn\cgt{\sum_{i=1}^4\bb_i + \a_+\sum_{i=1}^{N-1}m_i\be_i+ \a_-\sum_{i=1}^{N-1}n_i\be_i = 2\a_0\br}
is satisfied. \vs

We want to compute the  vacuum correlation function
\eqn\aba{G_4 = \langle D(\infty)\overline{\phi}(1)\phi(z)\overline{D}(0)\rangle}
where
\eqn\abaa{\phi \sim  (\L_+,0)~, \quad D \sim (0,\L_-)}
are in various representations discussed in the main text.
In the Coulomb gas formalism, the correlator \aba\ is especially manageable to calculate if either $\phi$ or $D$ is in either the $({\small\yng(1)},0)$ or $(0,{\small\yng(1)})$ representations, as is  the case throughout this paper. The reason is that these require a simple set of screening charges. Let us demonstrate with the correlator we will compute in the next subsection: taking $\phi \sim ({\small\yng(1)},0)$ which maps to the vertex operator
\eqn\cgz{\eqalign{({\small\yng(1)},0) &\mapsto V_{-\a_+ \o_1}~.}}
\aba\ can then be written as
\eqn\chd{G_{4} =  \oint dz_1\ldots dz_{N-1}\langle V_{2\a_0\hat{\rho}-\bb^*}(w)V_{-\a_+\bo_{N-1}}(1)V_{-\a_+\bo_1}(z)V_{\bb^*}(0)\prod_{i=1}^{N-1}V_{\a_+\be_i}(z_i)\rangle}
in the limit $w\rar\infty$ (and dropping overall factors of $w$). We have made a convenient choice of the charge vector for one of the defect operators: in particular, because $\o_1+\o_{N-1} = \sum_{i=1}^{N-1}\be_i$, the charge conservation condition \cgt\ merely requires a single screening charge $Q_+$ for each  simple root of SU(N). This is the case in \chd.
\subsec{SU(N) group theory}

We fix some of the notation and useful identities regarding group theoretic details of $SU(N)$. We denote the simple roots as ${\bf e}_i$ and the fundamental weights as $\o_i$. The Weyl vector is $\hat{\rho}=\sum_i \o_i$ and $A$ is the Cartan matrix.  An explicit decomposition of a fundamental weight as a linear combination of simple roots is
\eqn\aza{
\o_i=(\underbrace{{{N-i}\over{N}},\ldots,{{N-i}\over{N}}}_{i},\underbrace{-{{i}\over{N}},\ldots,-{{i}\over{N}}}_{N-i}   ).
}
Some of the identities used in this work are
\eqn\azb{
\eqalign{
\o_i\cdot \o_j &= {{(N-i)j}\over{N}} \quad , \quad i\geq j
\cr
{\bf e}_i \cdot {\bf e}_j &=A_{ij}
\cr
\o_i \cdot {\bf e}_j &=\delta_{ij}
\cr
\sum_i {\bf e}_i &=\o_1 +\o_{N-1}
\cr
{\bf e}_i \cdot \hat{\rho} &= 1~.
}
}

Introducing a weight vector $\Lambda_+=\sum_i d_i\o_i$, some useful identities are
\eqn\azc{
\eqalign{
\o_1 \cdot \Lambda_+ &= \sum\limits_{i=1}^{N-1} d_i -{{B}\over{N}}
\cr
\o_1 \cdot \hat{\rho} &={{N-1}\over{2}}
~,}
}
where $B$ is the total number of boxes in the Young diagram for $\Lambda_+$ .

\subsec{Supplement to subsection 4.2}

We follow the order of subsection 4.2, in which the scalar field is always in the defining representation, $\phi \sim({\small\yng(1)},0)$.
\vs
\noindent \bul {\it Defect in rectangular representation}\vs

First we present the result for the correlator
\eqn\aca{G_4 = \langle D_{n\times m}(\infty)\overline{\phi}(1)\phi(z)\overline{D}_{n\times m}(0)\rangle}
with  $D$ in a ``rectangular'' representation of $n$ columns each of $m$ boxes. This can be computed via \chd\ with $\bb^* = -\a_- n\o_{N-m}$.

Using \chii, the correlator is
\eqn\choa{\eqalign{&G_4= \Bigg|(1-z)^{2\a_+^2\over N}z^{-{nm\over N}}\oint dz_{N-m}(w-z_{N-m})^{4\a_+\a_0-n}z_{N-m}^n\cr
%&\cr
&\left[\oint dz_1\ldots dz_{N-m-1}(z-z_1)^{-2\a_+^2}\prod_{i=1}^{N-m-1}(w-z_i)^{4\a_+\a_0}(z_i-z_{i+1})^{-2\a_+^2}\right]\cr
&\Bigg[\oint dz_{N-m+1}\ldots dz_{N-1} (1-z_{N-1})^{-2\a_+^2}\left(\prod_{i=N-m+1}^{N-1}(w-z_i)^{4\a_0\a_+}\right)\cr
&\quad\quad\quad\quad\quad\quad\quad\quad\quad\quad\quad\quad\quad\quad\quad\quad \times \left(\prod_{i=N-m}^{N-2}(z_i-z_{i+1})^{-2\a_+^2}\right)\Bigg]\Bigg|^2~.}}
We have ignored overall factors involving $w$. The strategy is to first perform all integrals except the one over $z_{N-m}$, and what remains will be a hypergeometric integral that gives us our final answer. In doing so, the following is a useful result:
\eqn\chk{\oint dz_i\prod_{j=1}^3(z_i-w_j)^{2\b_j} \propto (w_1-w_2)^{\b_1+\b_2-\b_3} \times (cyclic)}
when $\sum_{j=1}^3\b_j=-2$. What remains after integration over all variables but $z_m$ is
\eqn\chsc{\eqalign{&G_4= \Big|(1-z)^{2\a_+^2\over N}z^{-{nm\over N}}\oint dy ~y^a(y-1)^{b}(y-z)^{c} \Big|^2}}
where
\eqn\chsd{a = n~, ~~ b =m-1-2\a_+^2m~, ~~ c = N-m-1 -2\a_+^2(N-m)~.}
To evaluate this, one must confront the issue of specifying the contour and imposing monodromy invariance. The correct result, following \DiFrancescoNK, turns out to be
\eqn\cht{|\oint dy y^a (y-1)^b(y-z)^c|^2 = {s(b)s(a+b+c)\over s(a+c)}|I_1(z)|^2+{s(a)s(c)\over s(a+c)}|I_2(z)|^2}
where $s(a) = \sin(\pi a)$, and the $I_i(z)$ are
\eqn\chu{\eqalign{I_1(z) &= {\Gamma(-a-b-c-1)\Gamma(b+1)\over \Gamma(-a-c)}{}_2F_1(-c,-a-b-c-1;-a-c;z)\cr
I_2(z) &= z^{1+a+b+c}{\Gamma(a+1)\Gamma(c+1)\over \Gamma(a+c+2)}{}_2F_1(-b,a+1;a+c+2;z)~.}}

Because $a\in \IZ$, the $I_2(z)$ piece drops out. Up to overall normalization, this gives the result \rda, \rdaa.

\vs
\noindent \bul {\it Defect in general  representation}\vs

We show now how to compute the correlator
\eqn\aya{ G_4=\langle
D_{(0,\L_-)}(\infty)\overline{\phi}(1)\phi(z)\overline{D}_{(0,\L_-)}(0)\rangle
 ,}
via {\chd\ }with $\bb = -\a_- \sum d_i\o_i$. Using \chii\ and some of the formulas in Appendix B.2, the integral expression of the correlator reads
\eqn\ayb{\eqalign{
G_4&=\Bigg|
(1-z)^{{{2\alpha_+^2}\over{N}}}z^{\sum\limits_{j=1}^{N-1}(1+d_j-2\alpha_+^2)-{{B}\over{N}}} \cr
&\times \oint dz_1 \ldots \oint dz_{N-1} {{
\prod\limits_{i=1}^{N-1}(w-z_i)^{d_i}z_i^{4\alpha_+^2-2-d_i}
}\over{
(z-z_1)^{2\alpha_+^2}\prod\limits_{i=1}^{N-2}(z_i-z_{i+1})^{2\alpha_+^2}(z_{N-1}-1)^{2\alpha_+^2}
}}
\Bigg|^2
 .}}

Let us focus on the integrals for now, one at a time, ignoring the
prefactors (and the antiholomorphic part) until the end. As we are
taking $w$ to infinity, we can also permanently ignore the factors
$(w-z_i)$ which do not affect the pole structure of the integrand.
We first need to specify the integration contour for the integral
over $z_1$. The integrand has branch points at $0,\infty, z_2,$ and
$z$. To ensure that we come back to the same sheet once we go around
the contour, we require no net winding number around any of the
branch points. There are two  independent choices and
they are known as Pochhammer contours. We will denote them ${\cal
P}(z,\infty)$ and ${\cal P}(0,z_2)$. These contours are only well
defined if the monodromies of the integral around both branch points
commute, and in that case they can be collapsed to a line interval
connecting the two branch points.  The result for the integral over
$z_1$ is parallel to what happened in equation {\cht}. The
contribution coming from ${\cal P}(z,\infty)$ is zero due to the
monodromy factor vanishing. The relevant contour in this case is the
line connecting  $0$ and $z_{i+1}$, so we need to evaluate
\eqn\ayc{
\int\limits_{0}^{z_2} dz_1
{
{
z_1^{4\alpha_+^2-2-d_1}
}\over{
(z-z_1)^{2\alpha_+^2}(z_1-z_2)^{2\alpha_+^2}
}
}
 .}

This is easily recognized as the integral representation of the hypergeometric function ${}_{2}F_{1}$. The result is
\eqn\ayd{
z^{-2\alpha_+^2}
{{\Gamma(1-2\alpha_+^2)\Gamma(4\alpha_+^2-1-d_1)}\over{\Gamma(2\alpha_+^2-d_1)}}
z_2^{2\alpha_+^2-1-d_1}{}_2F_{1}\left(  {{2\alpha_+^2,4\alpha_+^2-1-d_1}\atop{2\alpha_+^2-d_1}} \Big|{{z_2}\over{z}} \right)
 .}

Now we tevaluate the $z_2$ integral, ignoring the $z_2$-independent prefactor until the end. The $z_2$ dependent part of the integrand has poles at $0,\infty, z_3,$ and $z$, the last one coming from the hypergeometric function. In the main text, we provide a rigorous argument for why only one contour contributes. Given this, the integral over $z_2$ is
\eqn\ayf{
\int\limits_{0}^{z_3} dz_2
z_2^{6\alpha_+^2-3-d_1-d_2}(z_{2}-z_{3})^{-2\alpha_+^2}{}_2F_{1}\left(  {{2\alpha_+^2,4\alpha_+^2-1-d_1}\atop{2\alpha_+^2-d_1}} \Big|{{z_2}\over{z}} \right)
 ,}
which can be recognized as Euler's integral transform for generalized hypergeometric functions. The result is
\eqn\ayg{\eqalign{&
{{\Gamma(1-2\alpha_+^2)\Gamma(6\alpha_+^2-2-d_1-d_2)}\over{\Gamma(4\alpha_+^2-1-d_1-d_2)}}
z_3^{4\alpha_+^2-2-d_1-d_2}\cr&\times{}_3F_{2}\left(  {{2\alpha_+^2,4\alpha_+^2-1-d_1,6\alpha_+^2-2-d_1-d_2}\atop{2\alpha_+^2-d_1,4\alpha_+^2-2-d_1-d_2}} \Big|{{z_3}\over{z}} \right)
.}}

It continues to be the case for the remaining integrals over the other insertion points $z_i$, which are also Euler integral transforms of hypergeometric functions, that only one contour contributes. After doing the $N-1$ integrals and collecting all the gamma factors we are left with
\eqn\ayh{
z^{-2\alpha_+^2}\prod\limits_{k=1}^{N-1}\left({\Gamma(1-2\alpha_+^2)\Gamma(2\alpha_+^2-v_k)}\over{\Gamma(1-v_k)}\right)
{}_NF_{N-1}\left(  {{2\alpha_+^2,2\alpha_+^2\bf{1}-\bf{v}}\atop{\bf{1}-\bf{v}}}    \Big|{{1}\over{z}}\right)
 , }
where we have introduced a vector $\bf{v}$ with $N-1$ components $v_k=\sum\limits_{j=1}^{k}\left(1+d_j-2\alpha_+^2\right)$ in order to make the result appear more compact. Restoring the anti-holomorphic part and the $z$-dependent prefactors in \ayb, the result for the correlator reads
\eqn\axa{ \eqalign{ &\langle
D_{(0,\L_-)}(\infty)\overline{\phi}(1)\phi(z)\overline{D}_{(0,\L_-)}(0)\rangle
\propto\cr& \Big|C_{\L_-}
(1-z)^{{{2\alpha_+^2}\over{N}}}z^{v_{N-1}-{{B}\over{N}}-2\alpha_+^2}
{}_NF_{N-1}\left(
{{2\alpha_+^2,2\alpha_+^2\bf{1}-\bf{v}}\atop{\bf{1}-\bf{v}}}
\Big|{{1}\over{z}}\right) \Big|^2} , }
where
\eqn\axaa{ C_{\L_-} \equiv
\prod\limits_{k=1}^{N-1}\left({\Gamma(1-2\alpha_+^2)\Gamma(2\alpha_+^2-v_k)}\over{\Gamma(1-v_k)}\right)~.}

\subsec{Supplement to subsection 6.2}
Finally, we want to compute \aba\ with
 \eqn\sta{\phi \sim (\L_+,0) ~, ~~ D \sim(0,{\small\yng(1)})}
where the weight vector is $\L_+ = n\o_m$. This case is essentially identical to
the first computation in appendix B.3, except with the substitution
$\a_+\rar \a_-$ and permutations of coordinates among the operators.
Thus the Coulomb gas correlator we want to compute is
\eqn\nfa{G_4= \langle
V_{-\a_-\o_{N-1}}(w)V_{2\a_0\hat{\rho}-\bb^*}(1)V_{\bb^*}(z)V_{-\a_-\o_1}(0)\prod_{i=1}^{N-1}V_{\a_-e_i}(z_i)
\rangle}
with $\bb^* = -\a_+n\o_{N-m}$.

Performing the free field contractions and dropping the overall $w$ terms,
\eqn\nfh{\eqalign{&G_{4}=\cr
&\Bigg|(z-1)^{-\Delta_{n\times m}}z^{-{nm\over N}}\oint dz_1dz_{N-m}~z_1^{-2\a_-^2}(z_1-1)^{4\a_0\a_-}(z_{N-m}-1)^{4\a_0\a_--n}(z_{N-m}-z)^{n}\cr
&\times \Bigg[\Bigg(\prod_{i=2}^{N-m-1}\oint dz_i(z_i-1)^{4\a_0\a_-}(z_{i+1}-z_i)^{-2\a_-^2}\Bigg)(z_2-z_1)^{-2\a_-^2} \Bigg]\cr&
\times\Bigg[\Bigg(\prod_{i=N-m+1}^{N-2}\oint dz_i(z_i-z_{i-1})^{-2\a_-^2}(z_i-1)^{4\a_o\a_-}\Bigg)\times\cr& \quad (z_{N-1}-w)^{-2\a_-^2}(z_{N-1}-1)^{4\a_0\a_-}(z_{N-1}-z_{N-2})^{-2\a_-^2}\Bigg]\Bigg|^2}}
where
\eqn\stda{\Delta_{n\times m} = -nm(N-m) + 2\a_+^2\left({nm(N-m)(n+2N)\over N}\right)~.}
We have arranged the terms to simplify the integrations. First, perform the $m-1$ integrals over $i=N-m+1\ldots N-1$. Then do the $N-m-2$ integrals over $i=2\ldots N-m-1$. Finally, do the integral over $z_1$. What remains (after throwing out various factors of $w$) is an integral of the form \cht\ with
\eqn\nfhb{a=-2\a_-^2-2\a_0\a_-(N-m-1)~, ~~ b = 2\a_0\a_-N-n~, ~~ c=n~.}
Again, only one integration contour contributes because $c\in \IZ$. Using \cgp\ to substitute $\a_0 = \a_++\a_-$ and plugging into \chu, the result is \stb\ and \stc.

\listrefs
\end

%% file: youngtab.tex
%%
%% This is file `youngtab.tex',
%% (manually) generated from `youngtab.sty'
%% (For use with TeX)
%%
%% The original source files were:
%%
%% youngtab.dtx  (with options: `package')
%% 
%% Copyright (C) 1996,98,99 Volker B"orchers and Stefan Gieseke,
%% This program can be redistributed and/or modified under the terms
%% of the LaTeX Project Public License Distributed from CTAN
%% archives in directory macros/latex/base/lppl.txt; either
%% version 1 of the License, or any later version.
\catcode`\@11\relax
\newif\ify@autoscale \y@autoscaletrue \def\Yautoscale#1{\ifnum #1=0
  \y@autoscalefalse\else\y@autoscaletrue\fi}
\newdimen\y@b@xdim
\newdimen\y@boxdim \y@boxdim=13pt
\def\Yboxdim#1{\y@autoscalefalse\y@boxdim=#1}
\newdimen\y@linethick    \y@linethick=.3pt
\def\Ylinethick#1{\y@linethick=#1}
\newskip\y@interspace \y@interspace=0ex plus 0.3ex
\def\Yinterspace#1{\y@interspace=#1}
\newif\ify@vcenter   \y@vcenterfalse
\def\Yvcentermath#1{\ifnum #1=0 \y@vcenterfalse\else\y@vcentertrue\fi}
\newif\ify@stdtext   \y@stdtextfalse
\def\Ystdtext#1{\ifnum #1=0 \y@stdtextfalse\else\y@stdtexttrue\fi}
\newif\ify@enable@skew   \y@enable@skewfalse
%% To use skew tableaux, define a macro \enableskew
%% right before loading this file: \def\enableskew{1}
\expandafter\ifx\csname enableskew\endcsname\relax
 \y@enable@skewfalse \else \y@enable@skewtrue\fi
%% \DeclareOption{noautoscale}{\y@autoscalefalse}
%% \DeclareOption{vcentermath}{\y@vcentertrue}
%% \DeclareOption{stdtext}{\y@vcentertrue}
%% \DeclareOption{enableskew}{\y@enable@skewtrue}
%% \DeclareOption*{\PackageWarning{youngtab}{%
%%     Unknown option `\CurrentOption' (Known:\MessageBreak
%%     `vcentermath', `noautoscale', `stdtext', `enableskew'.)}}
%% \ProcessOptions\relax
\def\y@vr{\vrule height0.8\y@b@xdim width\y@linethick depth 0.2\y@b@xdim}
\def\y@emptybox{\y@vr\hbox to \y@b@xdim{\hfil}}
\ify@enable@skew
 \def\y@abcbox#1{\if :#1\else
   \y@vr\hbox to \y@b@xdim{\hfil#1\hfil}\fi}
 \def\y@mathabcbox#1{\if :#1\else
   \y@vr\hbox to \y@b@xdim{\hfil$#1$\hfil}\fi}
\else
 \def\y@abcbox#1{\y@vr\hbox to \y@b@xdim{\hfil#1\hfil}}
 \def\y@mathabcbox#1{\y@vr\hbox to \y@b@xdim{\hfil$#1$\hfil}}
\fi
\def\y@setdim{%
  \ify@autoscale%
   \ifvoid1\else\typeout{Package youngtab: box1 not free! Expect an
     error!}\fi%
   \setbox1=\hbox{A}\y@b@xdim=1.6\ht1 \setbox1=\hbox{}\box1%
  \else\y@b@xdim=\y@boxdim \advance\y@b@xdim by -2\y@linethick
  \fi}
\newcount\y@counter
\newif\ify@islastarg
\def\y@lastargtest#1,#2 {\if\space #2 \y@islastargtrue
  \else\y@islastargfalse\fi}
\def\y@emptyboxes#1{\y@counter=#1\loop\ifnum\y@counter>0
  \advance\y@counter by -1 \y@emptybox\repeat}
\def\y@nelineemptyboxes#1{%
  \vbox{%
    \hrule height\y@linethick%
    \hbox{\y@emptyboxes{#1}\y@vr}
    \hrule height\y@linethick}\vskip-\y@linethick}
\def\yng(#1){%
  \y@setdim%
  \hskip\y@interspace%
  \ifmmode\ify@vcenter\vcenter\fi\fi{%
  \y@lastargtest#1,
  \vbox{\offinterlineskip
    \ify@islastarg
     \y@nelineemptyboxes{#1}
    \else
     \y@ungempty(#1)
    \fi}}\hskip\y@interspace}
\def\y@ungempty(#1,#2){%
  \y@nelineemptyboxes{#1}
  \y@lastargtest#2,
  \ify@islastarg
   \y@nelineemptyboxes{#2}
  \else
   \y@ungempty(#2)
  \fi}
\def\y@nelettertest#1#2. {\if\space #2 \y@islastargtrue
  \else\y@islastargfalse\fi}
\def\y@abcboxes#1#2.{%
  \ify@stdtext\y@abcbox#1\else\y@mathabcbox#1\fi%
  \y@nelettertest #2.
  \ify@islastarg\unskip%
   \ify@stdtext\y@abcbox{#2}\else\y@mathabcbox{#2}\fi%
  \else\y@abcboxes#2.\fi}
 \newdimen\y@full@b@xdim
 \newcount\y@m@veright@cnt
\ify@enable@skew
 \def\y@get@m@veright@cnt#1#2.{%
   \if :#1 \advance\y@m@veright@cnt by 1\y@get@m@veright@cnt#2.\fi}
 \let\y@setdim@=\y@setdim
 \def\y@setdim{%
   \y@setdim@ \y@full@b@xdim=\y@b@xdim
   \advance\y@full@b@xdim by 1\y@linethick}
 \def\y@m@veright@ifskew#1{
   \y@m@veright@cnt=0 \y@get@m@veright@cnt#1.
   \moveright \y@m@veright@cnt\y@full@b@xdim}
\else
 \def\y@m@veright@ifskew#1{}
\fi
\def\y@nelineabcboxes#1{%
  \y@nelettertest #1.
  \ify@islastarg
   \y@m@veright@ifskew{#1}
    \vbox{
      \hrule height\y@linethick%
      \hbox{\ify@stdtext\y@abcbox#1\else\y@mathabcbox#1\fi\y@vr}
      \hrule height\y@linethick}\vskip-\y@linethick
  \else
   \y@m@veright@ifskew{#1}
    \vbox{
      \hrule height\y@linethick%
      \hbox{\y@abcboxes #1.\y@vr}%
      \hrule height\y@linethick}\vskip-\y@linethick
  \fi}
\def\young(#1){%
  \y@setdim%
  \hskip\y@interspace%
  \y@lastargtest#1,
  \ifmmode\ify@vcenter\vcenter\fi\fi{%
  \vbox{\offinterlineskip
    \ify@islastarg\y@nelineabcboxes{#1}%
    \else\y@ungabc(#1)%
    \fi}}\hskip\y@interspace}
\def\y@ungabc(#1,#2){%
  \y@nelineabcboxes{#1}%
  \y@lastargtest#2,
  \ify@islastarg\y@nelineabcboxes{#2}%
  \else\y@ungabc(#2)%
  \fi}
\catcode`\@12\relax
 
%%
%% End of file `youngtab.tex'.